\documentclass[pra,twocolumn,showpacs,floatfix]{revtex4}
\usepackage[pdftex]{graphicx}
\usepackage{amsmath,amsfonts,amsbsy,amssymb,amsthm}
\usepackage{mathrsfs,bbm}
\newcommand{\ket}[1]{\lvert #1 \rangle}
\newcommand{\bra}[1]{\langle #1 \rvert}
\newcommand{\ketbra}[2]{\ket{#1}\bra{#2}}
\newcommand{\braket}[2]{\langle #1 \rvert #2 \rangle}
\newcommand{\Tr}[1]{\operatorname{Tr}\bigl[#1\bigr]}

\newcommand{\modsq}[1]{\lvert #1 \rvert^2}
\newcommand{\abs}[1]{\lvert #1 \rvert}

\newcommand{\expect}[1]{\langle #1 \rangle}
\newcommand{\partialD}[2]{\frac{\partial #1}{\partial #2}}

\newcommand{\Fx}{\hat{F}_{\mathrm{x}}}
\newcommand{\Fy}{\hat{F}_{\mathrm{y}}}
\newcommand{\Fz}{\hat{F}_{\mathrm{z}}}
\newcommand{\Fk}{\hat{F}_{k}}
\newcommand{\Fpx}{\hat{F}_{+,x}}
\newcommand{\Fmx}{\hat{F}_{-,x}}
\newcommand{\St}{{\hat{S}_t}}
\newcommand{\Yt}{\hat{Y}_{\theta_t}}
\newcommand{\Sqt}{\hat{S}_{\xi_t}}

\def\sHOmegaZero{\hat{s}^{\phantom{\dagger}}_{0,\omega}}

\def\sHOmegaX{\hat{s}^{\phantom{\dagger}}_{\mathrm{x},\omega}}

\def\sHOmegaY{\hat{s}^{\phantom{\dagger}}_{\mathrm{y},\omega}}

\def\sHOmegaZ{\hat{s}^{\phantom{\dagger}}_{\mathrm{z},\omega}}

\def\aDOmegaX{\hat{a}^\dagger_{\mathrm{x},\omega}}
\def\aHOmegaX{\hat{a}^{\phantom{\dagger}}_{\mathrm{x},\omega}}
\def\aDOmegaY{\hat{a}^\dagger_{\mathrm{y},\omega}}
\def\aHOmegaY{\hat{a}^{\phantom{\dagger}}_{\mathrm{y},\omega}}

\def\aDOmegaP{\hat{a}^\dagger_{\mathrm{+},\omega}}
\def\aHOmegaP{\hat{a}^{\phantom{\dagger}}_{\mathrm{+},\omega}}
\def\aDOmegaM{\hat{a}^\dagger_{\mathrm{-},\omega}}
\def\aHOmegaM{\hat{a}^{\phantom{\dagger}}_{\mathrm{-},\omega}}

\begin{document}
\title{Magnetometry via a double-pass continuous quantum measurement of atomic spin}
\author{Bradley A. Chase}
\email{bchase@unm.edu}
\author{Ben Q. Baragiola}
\author{Heather L. Partner}
\author{Brigette D. Black}
\author{JM Geremia}
\email{jgeremia@unm.edu}
\affiliation{Quantum Measurement \& Control Group, Department of Physics \& Astronomy, The University of New Mexico, Albuquerque, New Mexico 87131 USA}
\date{\today}
\begin{abstract}
We argue that it is possible in principle to reduce the uncertainty of an atomic magnetometer by double-passing a far-detuned laser field through the atomic sample as it undergoes Larmor precession.   Numerical simulations of the quantum Fisher information suggest that, despite the lack of explicit multi-body coupling terms in the system's magnetic Hamiltonian, the parameter estimation uncertainty in such a physical setup scales better than the conventional Heisenberg uncertainty limit over a specified but arbitrary range of particle number $N$.  Using the methods of quantum stochastic calculus and filtering theory, we demonstrate numerically an explicit parameter estimator (called a quantum particle filter) whose observed scaling follows that of our calculated quantum Fisher information.  Moreover, the quantum particle filter quantitatively surpasses the uncertainty limit calculated from the quantum Cram\'{e}r-Rao inequality based on a magnetic coupling Hamiltonian with only single-body operators.  We also show that a quantum Kalman filter is insufficient to obtain super-Heisenberg scaling, and present evidence that such scaling necessitates going beyond the manifold of Gaussian atomic states.
 
\end{abstract}
\maketitle
\section{Introduction}

It is well-appreciated in physics that the properties of a field must often be determined indirectly, such as by observing the effect of the field on a test particle.  Take magnetometry for example:  the strength of a magnetic field might be inferred by observing Larmor precession in a spin-polarized atomic sample \cite{Budker:2002a} and estimating the field strength $B$ from the precession rate.  Inherent in this process is the fact that the atomic spin must be measured to determine the extent of the magnetically-induced dynamics.  For very precise measurements, uncertainty $\delta \tilde{B}$ in the estimated value $\tilde{B}$ of the field is dominated by quantum fluctuations in the observations performed on the atomic sample.   The results presented here fall under the umbrella of quantum parameter estimation theory \cite{Helstrom:1976a,Braunstein:1994a}, where the objective is to work within the rules of quantum mechanics to minimize, as much as possible, the propagation of this quantum uncertainty into the determination of metrological quantities, like $B$.

Given, for instance, a $y$-axis magnetic field $\mathbf{B} = B\,  \vec{\mathbf{y}}$, an atomic sample couples to $B$ via the magnetic dipole Hamiltonian
\begin{equation} \label{Equation::HLarmor}
	\hat{H} = - \hbar \gamma B \Fy,
\end{equation}
where $\gamma$ is the atomic gyromagnetic ratio and $\hat{F}_i =\sum_{j=1}^{N_A} \hat{f}_i^{(j)}$ ($i=\mathrm{x}, \mathrm{y}, \mathrm{z}$) are the collective spin operators obtained from a symmetric sum over $N_A$ identical spin-$f$ atoms.  If the atoms are initially polarized along the $x$-axis, the Larmor dynamics and thus $B$ can be inferred by observing the $z$-component of the atomic spin $F_\mathrm{z}$ \cite{Budker:2002a,Romalis:2003a,Geremia:2003a,Molmer:2004a}.

It is possible to place an information-theoretic lower bound on the units-corrected mean-square deviation of any magnetic field estimator $\tilde{B}$ from the actual value of the field $B$,
\begin{equation} \label{Equation::CRError}
    \delta \tilde{B} = \left\langle\left(\frac{\tilde{B}}
    {\abs{d\expect{\tilde{B}}/dB}} - B\right)^2 \right\rangle^{1/2} .
\end{equation}
via the quantum Cram\'{e}r-Rao inequality (as described in Section \ref{section:cr}) \cite{Helstrom:1976a,Holevo:1982a,Braunstein:1994a,Braunstein:1995a}.  The behavior of the estimator uncertainty with the number of atoms $N_A$ depends on the characteristics (e.g., separable, entangled, etc.) of the quantum states used to compute the expectation value in Eq.\ (\ref{Equation::CRError}) as well as the nature of the induced dynamics \cite{Boixo:2007a}.  If one does not permit quantum entanglement between the different atoms in the probe, it can be shown that the optimal parameter resolution obtained from Eq.\ (\ref{Equation::HLarmor}), evolved for a time $t$, is given by the so-called shotnoise uncertainty \cite{Budker:2002a,Geremia:2003a}
\begin{equation} \label{Equation::DeltaBSN}
	\delta \tilde{B}_\mathrm{SN}(t) = \frac{1} { \gamma t \sqrt{2 F }} ,
\end{equation}
whose characteristic $1/\sqrt{N_A}$ scaling is a byproduct of the projection noise $\langle \Delta \Fz\rangle =\sqrt{F/2}$ for a spin coherent state \cite{Wineland:1994a} (here $F = f N_A$ for a sample of $N_A$ atoms each with total spin quantum number $f$).  It was believed for some time that the fundamental limit to parameter estimation, even when exploiting arbitrary entanglement between atoms in the probe, offers only a quadratic improvement
\begin{equation} \label{Equation::DeltaBHL}
	\delta \tilde{B}_\mathrm{HL}( t ) = \frac{\alpha} { \gamma t F} ,
\end{equation}
up to an implementation-dependent constant $\alpha$.  Eq.\ (\ref{Equation::DeltaBHL}) has traditionally been called the Heisenberg uncertainty scaling, and it can be achieved in principle for various spin resonance metrology problems \cite{Wineland:1994a}, including magnetometry \cite{Geremia:2003a}.  For an ensemble of $N_A$ spin-1/2 particles prepared into the initial cat-state $( \ket{\uparrow\uparrow \cdots \uparrow} + \ket{\downarrow\downarrow\cdots\downarrow})/\sqrt{2}$, the uncertainty scaling is given by $1/\gamma t N_A$ and is sometimes called the \textit{Heisenberg Limit}. 

Recently, however, it was shown that $1/N_A$ scaling can be surpassed \cite{Boixo:2007a} by extending the linear coupling that underlies Eq.\ (\ref{Equation::HLarmor}) to allow for multi-body collective interactions \cite{Boixo:2007a,Rey:2007a}.  Were one to engineer a probe Hamiltonian where $B$ multiplies $k$-body probe operators, such as $\Fy^k$, then the quantum Cramer-Rao bound \cite{Braunstein:1994a} indicates that the optimal estimation uncertainty would scale more favorably as $\Delta B_k \sim 1/N_A^k$ \cite{Boixo:2007a}.   Unfortunately, metrological coupling Hamiltonians are rarely up to us--- they come from nature, like the Zeeman interaction--- suggesting that one is stuck with a given uncertainty scaling without changing the fundamental structure of Eq.\ (\ref{Equation::HLarmor}).  Furthermore, it was shown in Ref.\ \cite{Boixo:2007a} that the addition of an auxiliary parameter-independent Hamiltonian $\hat{H}_1(t)$ such that
\begin{equation} \label{Equation::Haux}
	\hat{H} = - \hbar \gamma B \hat{F}_\mathrm{y} + \hat{H}_1(t)
\end{equation}
does not change the scaling of the parameter uncertainty for any choice of $\hat{H}_1(t)$.

At the same time, however, it should be well-appreciated that the dynamics one encounters in any actual physical setting are \textit{effective dynamics}.  Indeed, even the hyperfine Zeeman Hamiltonian Eq.\ (\ref{Equation::HLarmor}) is an effective description at some level.  This begs the question as to whether one can utilize an auxiliary system to induce effective dynamics that improve the uncertainty scaling in quantum parameter estimation by going outside the structure of Eq.\ (\ref{Equation::Haux}).  The purpose of this paper is to provide some direct evidence that doing so is possible.

\begin{figure}[t]
\begin{center}
\includegraphics{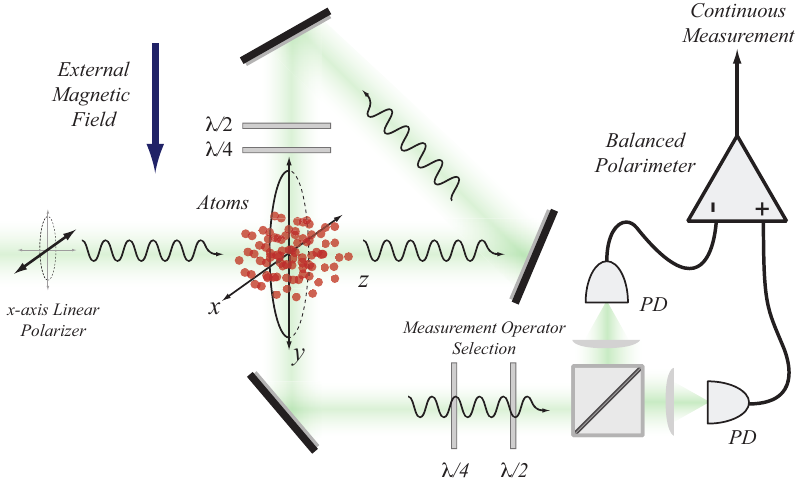}
\end{center}
\vspace{-4mm}
\caption{(color online) Schematic of a broadband atomic magnetometer based on continuous observation of a polarized optical probe field double-passed through the atomic sample.  \label{Figure::Schematic}}
\end{figure}

We study effective nonlinear couplings generated by double-passing an optical field through an atomic sample (q.v. Figure \ref{Figure::Schematic}) \cite{Sherson:2006a,Sarma:2008a,Muschik:2006a}.  Continuous measurement of the scattered field then allows for the estimation of $\Fz$ and by extension, the magnetic field.  Building on the quantum stochastic calculus approach in \cite{Sarma:2008a}, we derive the quantum filtering equations for estimating the state of the atomic sample.  Although the effective dynamics are no longer described by a Hamiltonian, we perform numerical calculations of the quantum Fisher information to obtain a theoretical lower bound on the uncertainty scaling of an optimal magnetic field estimator \cite{Braunstein:1994a}.  Our simulations suggest that for certain parameter regimes, the double-pass system's sensitivity to magnetic fields scales better than that of a comparable single-pass system and what would be computed by applying the methods of Ref.\ \cite{Boixo:2007a} to Eq.\ (\ref{Equation::Haux}).  We also conduct direct simulations of magnetic field estimation for our system using a general nonlinear parameter estimator, called a quantum particle filter \cite{Chase:2008a}, as further evidence for the improved uncertainty scaling provided by our proposed magnetometer.

Unfortunately our results are somewhat muted by the fact that we have, despite our best efforts, not yet found a parameter estimator whose uncertainty scaling can be shown analytically to outperform the conventional Heisenberg limit.  In particular, we show that improved scaling is not achieved by a quantum Kalman filter \cite{Belavkin:1999a,Kalman:1960a,Kalman:1961a}, as such a filter is only suitable for estimating magnetic fields in the linear small-angle regime and where the state is Gaussian and the dynamics are well approximated by a low order Holstein-Primakoff expansion \cite{Holstein:1940a,Geremia:2003a}.   Although Kalman filters have had success in describing the single-pass system \cite{Geremia:2003a}, simulations suggest the Gaussian and small-angle approximations break down precisely when exact simulations of the double-pass system show improved sensitivity.  For pedagogical purposes, we detail the derivation of such linear-Gaussian filters using the method of projection filtering \cite{vanHandel:2005b,Mabuchi:2008a}.  Doing so allows us to observe directly the limitations that arise when imposing the small-angle and Gaussian assumptions, and it also provides a framework for the future development of more sophisticated filters.    As a consequence, our analysis here is restricted to values of the total atomic angular momentum $F$ that is small compared to experimentally relevant values.   It remains an important problem to develop nonlinear parameter estimation methods suitable for implementation in an actual laboratory setting.

The remainder of the paper is organized as follows.  In Section \ref{sec:continuous_measurement_of_double_pass_system}, we detail the magnetometer setup and its corresponding quantum stochastic description.  We also present the quantum filtering equation which describes continuous measurement of the double-passed atomic sample.  Using this model, we present a numerical analysis of the quantum Cram\'{e}r-Rao bound in Section \ref{section:cr}) and observe an improved uncertainty scaling relative to the shotnoise and Heisenberg scalings.  In Section \ref{sec:magnetic_field_estimators}, we use the double-pass quantum filter to develop quantum particle filters and quantum Kalman filters suitable for estimating an unknown magnetic field.  In Section \ref{sec:simulations}, we present simulations of the particle filter magnetic field estimator and discuss the evidence for improved magnetic field sensitivity in the double-pass system.  We conclude in Section \ref{sec:conclusion}.

\section{Continuous measurement of the double-pass system} 
\label{sec:continuous_measurement_of_double_pass_system}

Consider the schematic in Fig.\ \ref{Figure::Schematic}.  The objective of this apparatus is to estimate the strength of a magnetic field oriented along the laboratory $y$-axis by observing the effect of that field on the spin state of the atomic sample.  Like most atomic magnetometer configurations, our procedure relies upon Larmor precession and uses a far-detuned laser probe to observe the spin angular momentum of the atomic sample.  Unlike conventional atomic magnetometer configurations, however, the probe laser is routed in such a way that it passes through the atomic sample twice prior to detection \cite{Sherson:2006a,Sarma:2008a,Muschik:2006a}. 

Qualitatively, the magnetometer operates as follows.  The incoming probe field propagates initially along the atomic $z$-axis and is linearly polarized.  As a result of the atomic polarizability of the atoms, the probe laser polarization acquires a Faraday rotation proportional to the $z$-component of the collective atomic spin.  Two folding mirrors are then used to direct the forward scattered probe field to pass through the atomic sample a second time, now propagating along the atomic $y$-axis.  Prior to its second interaction with the atoms, polarization optics convert the initial Faraday rotation into ellipticity.  Thus on the second pass, the atoms perceive the optical helicity as a fictitious $y$-axis magnetic field acting in addition to the real field $B$, providing a positive feedback effect modulated by the strength of $B$.  That is, the total Larmor precession of the spins is enhanced by an amount which depends on the true strength of $B$.  The twice forward-scattered optical field is then measured in such a way that is sensitive only to the Faraday rotation induced by the first pass atom-field interaction.

\subsection{Quantum Stochastic Model}

When the collective spin angular momentum of a multilevel atomic system interacts dispersively with a traveling wave laser field with wavevector $\mathbf{k}$, the atomic spin couples to the two polarization modes of the electromagnetic field transverse to $\mathbf{k}$.   These polarization modes can can be viewed as a Schwinger-Bose field that when quantized in terms of a plane-wave mode decomposition yields the familiar Stokes operators:
\begin{eqnarray}
 	\sHOmegaZero & = &  +\frac{1}{2} \left(
				\aDOmegaX \aHOmegaX + \aDOmegaY \aHOmegaY 
			\right) \\ 
			& = & + \frac{1}{2} \left(
				\aDOmegaP \aHOmegaP + \aDOmegaM \aHOmegaM
			\right) \nonumber \\
	 \sHOmegaX & = & +\frac{1}{2} \left(
			\aDOmegaY \aHOmegaY - \aDOmegaX \aHOmegaX	 
	 	\right) \\
				& = & + \frac{1}{2} \left(
				\aDOmegaP \aHOmegaM + \aDOmegaM \aHOmegaP 	
			\right) \nonumber \\
	 \sHOmegaY & = & - \frac{1}{2} \left(
	 		\aDOmegaY \aHOmegaX + \aDOmegaX \aHOmegaY
	 	\right) \\
		& = & - \frac{i}{2} \left(
			\aDOmegaP \aHOmegaM -  \aDOmegaM \aHOmegaP
		\right) \nonumber \\
	 \sHOmegaZ & = & + \frac{i}{2} \left( 
	 		\aDOmegaY \aHOmegaX - \aDOmegaX \aHOmegaY
	 	\right) \\
		& = & + \frac{1}{2} \left(
			\aDOmegaP \aHOmegaP - \aDOmegaM \aHOmegaM
		\right). \nonumber		
 \end{eqnarray} 
Here, we have expressed the Stokes operators in terms of the Schr\"{o}dinger-picture field annihilation operators, $\aHOmegaX$ and $\aHOmegaY$, for the plane-wave modes with frequency $\omega$ and linear polarization along the x- and y-axes, respectively, as well as their corresponding transformations into the spherical polarization basis.  

In developing a physical model for the atom-field interaction in Fig.\ \ref{Figure::Schematic}, it is convenient to transform from a plane-wave mode decomposition of the electromagetic field to operators that are labeled by time.  Towards this end, we define the time-domain Schwinger boson annihilation operator as the operator distribution
\begin{equation}
	\hat{s}_t = \frac{1}{2} \int_{-\infty}^{+\infty}  g(\omega) 
		\, \aDOmegaX \aHOmegaY e^{i \omega t} d \omega,
\end{equation}
where $g(\omega)$ is a form factor.  This definition permits us to express the Stokes operators as
\begin{equation}
	\hat{s}_{\mathrm{z},t} =  i 
		\left(  \hat{s}^\dagger_t - \hat{s}^{\phantom{\dagger}}_t \right) \quad\text{and}\quad
	\hat{s}_{\mathrm{y},t}  =  - \left( \hat{s}^{\phantom{\dagger}}_t + \hat{s}^\dagger_t  \right),
\end{equation}
which should be reminiscent of quadrature operators and also places the field operators in a form that is directly in line with the standard nomenclature adopted in the field of quantum stochastic calculus. 

With a suitable orientation of the polarization optics ($\lambda/2$ and $\lambda/4$) in Fig.\ \ref{Figure::Schematic}, the interaction Hamiltonians for each pass of the probe light through the sample are then
\begin{eqnarray} 
		\hat{H}_t^{(1)}  & = & + \hbar \mu \Fz \hat{s}_{\mathrm{z},t}
			=  + i\hbar\mu\Fz \left( \hat{s}_t^{\dagger} - \hat{s}_t^{\phantom{\dagger}} \right)
		\label{Equation::H1} \\
		\hat{H}_t^{(2)}  &= &  +\hbar \kappa \Fy \hat{s}_{\mathrm{y},t} =  -\hbar\kappa \hat{F}_\mathrm{y} \left( \hat{s}_t^{\phantom{\dagger}} + \hat{s}_t^\dagger \right),
		\label{Equation::H2}
\end{eqnarray}
respectively, where the coupling strengths $\mu,\kappa$ arise from the atomic-polarizibility level structure for the particular atoms being used \cite{Jessen:2004a,Geremia:2006b}.  Note that in developing these Hamiltonians, which are of the standard atomic polarizability form, it was assumed the laser frequency $\omega_l$ is far detuned from any atomic resonance so that rank-two spherical tensor interactions can be neglected \cite{Jessen:2004a,Geremia:2006b}.  In practice, the validity of such an assumption can depend heavily on the choice of atomic level structure and experimental parameters such as the intensity and detuning of the probe laser field.

In addition to specifying the Hamiltonians for the two atom-field interactions, it is also necessary to stipulate the measurement to be performed on the probe laser.  Since we expect that the amount of Larmor precession (possibly augmented by the addition of the double-passed probe field) will cary information about the magnetic field strength $B$, we must choose the measured field operator $\hat{Z}_t$ appropriately.   Since the magnetic field drives rotations about the atomic $y$-axis, it is the $z$-component of the atomic spin that indicate such a rotation.  From the form of the first-pass interaction Hamiltonian $\hat{H}^{(1)}_t$, we see that the $z$-component of the atomic spin couples to dynamics generated by the field operator $\hat{s}_{\mathrm{z},t} = i ( \hat{s}^\dagger_t - \hat{s}^{\phantom{\dagger}}_t)$.  The affect of such a coupling is then observed by measuring the orthogonal quadrature, indicating that the appropriate polarization measurement should be $\hat{s}_{\mathrm{y,t}}$.

\subsubsection{The Stochastic Propagator}

Analyzing the two individual interactions $\hat{H}_1$ and $\hat{H}_2$ is a well studied problem in continuous measurement theory as a weak-coupling limit \cite{Accardi:2002a,Accardi:1990a,Gough:2005a,vanHandel:2005a}.  Heuristically, we are interested in the long time cumulative effects of the fast field dynamics relative to the slow atomic dynamics.  Much like the classical central limit theorem, the above references show that the cumulative effect of the field converges to a white noise process, so that the overall interaction-picture evolution of the joint atom-field system is well-described by the following quantum stochastic differential equation (QSDE) propagators
\begin{align}
	d\hat{U}_t^{(1)} &= \left\{\sqrt{m}\Fz (dS_t^\dagger - dS_t) 
			- \frac{1}{2}m\Fz^2dt 
			- \frac{i}{\hbar} \hat{H} dt \right\}U_t^{(1)} \label{Equation::dU1}\\
	d\hat{U}_t^{(2)} &= \left\{i\sqrt{k}\Fy(dS_t + dS^{\dagger}_t) 
			- \frac{1}{2}k\Fy^2dt
			- \frac{i}{\hbar} \hat{H}dt \right\}U_t^{(2)} \label{Equation::dU2}
\end{align}
As discussed in \cite{Accardi:2002a}, the rates $m=2\pi |\mu g(\omega_l)|^2$ and $k=2\pi |\kappa g(\omega_l)|^2$ are evaluated in terms of the original coupling strengths and form factor at the drive laser frequency $\omega_l$,
 $\hat{H}$ is an arbitrary atomic Hamiltonian and $d\St^\dagger$ and $d \St$ are delta-correlated noise operators derived from the quantum Brownian motion
\begin{equation}
	\St = \int_0^t \hat{s}_u du.
\end{equation}
The noise terms satisfy the quantum It\^o rules: $d \St d \St^\dagger = dt$ and $d \St^\dagger d \St = d \St^2 = (d \St^\dagger)^2 = 0$, and can be viewed heuristically as a consequence of vacuum fluctuations in the probe field.  Note that these are field operators and are defined so as to commute with all operators $\hat{X}_s$ for $s \leq t $.

To obtain a single weak-coupling limit for the double-pass interaction, we combine the separate equations of motion for the two propagators into a single weak-couping limit as follows.  First, write the two single-pass evolutions in terms of the generators of the dynamics
\begin{equation}
	d\hat{U}_t^{(1)} = \hat{a}_t U_t^{(1)}, \quad \mathrm{and},
	\quad
	d\hat{U}_t^{(2)} = \hat{b}_t U_t^{(1)}
\end{equation}
and then expand the differential $d\hat{U}_t$ of the combined propagator
\begin{eqnarray}
	d \hat{U}_{t+\delta t}  & = & ( \hat{1} +\hat{b}_t ) ( \hat{1} + \hat{a}_t ) \hat{U}_t \\
	& = &  \hat{U}_t + \left( \hat{a} + \hat{b} + \hat{b} \hat{a} \right) \hat{U}_t
\end{eqnarray}
such that the combined propagator $d \hat{U}_t = \hat{U}_{t+\delta t} - \hat{U}_t$ then satisfies
\begin{equation}
	d\hat{U}_t = \left( \hat{a} + \hat{b} + \hat{b} \hat{a} \right) \hat{U}_t \,.
\end{equation}
After evaluating the combined evolution for the propagators in Eqs.\ (\ref{Equation::dU1}) and (\ref{Equation::dU2}) in light of the quantum It\^{o} rules, we find that the single weak-coupling limit propagator satisfies
\begin{multline}
	d\hat{U}_t=   \left[i\sqrt{km}\Fy\Fz dt 
		- \frac{1}{2}m\Fz^2 dt - \frac{1}{2}k\Fy^2dt  - \frac{2 i}{\hbar} \hat{H} dt \right. \\
		 \left. 
			+ \sqrt{m}\Fz(d\St^{\dagger} - d\St)
			+ i \sqrt{k}\Fy(d\St^{\dagger} + d\St) \right]\hat{U}_t. 
\end{multline}
Observe that as a result of the manner in which the combined weak-coupling limit was taken, the Hamiltonian term has the property that rates which appear in it differ by a factor of two from those that would be expected from a single weak-coupling limit.  This factor of two is essentially the rescaling of time units that arises from aggregating two sequential weak-coupling limits as a single differential process.  To retain consistency with the original definition of the frequencies that appear in the parameter-coupling Hamiltonian, it is essential to rescale time units such that frequencies in the parameter-coupling Hamiltonian are as expected.  Doing so is accomplished by reversing the effective $2 dt  \rightarrow dt$ transformation that occurred in the derivation, and thus dividing all rates by two to give
\begin{multline} \label{eq:double_pass_prop}
	d\hat{U}_t =  \left[i\sqrt{KM}\Fy\Fz dt 
		- \frac{1}{2}M\Fz^2 dt - \frac{1}{2}K\Fy^2dt  - \frac{i}{\hbar} \hat{H} dt \right. \\
		 \left. 
			+ \sqrt{M}\Fz(d\St^{\dagger} - d\St)
			+ i \sqrt{K}\Fy(d\St^{\dagger} + d\St) \right]\hat{U}_t
\end{multline}
where $M = m / 2$ and $K= k/2$.  We note that this final result agrees with the propagator obtained by Sarma~et.~al \cite{Sarma:2008a} who also derived the quantum stochastic propagator of this system in order to characterize the generation of polarization and spin squeezing as suggested by Sherson and M{\o}lmer \cite{Sherson:2006a} and Muschik et. al \cite{Muschik:2006a}.

\subsection{Double-Pass Quantum Filter} \label{app:quantum_filter}

Continuous polarimetry measurement of the transmitted probe field is well described by a quantum stochastic calculus model \cite{Bouten:2007a} which in turn allows for statistical inference or quantum filtering of the state of the atomic sample \cite{vanHandel:2005a,Bouten:2006a}.  We now derive the quantum filtering equation corresponding to the propagator in  Eq.\ (\ref{eq:double_pass_prop}) and continuous observation of the forward-scattered probe polarization.  Our derivation follows the general approach developed in Section 4 of Ref.\ \cite{vanHandel:2005a}.  In this condensed presentation, we gloss over some of the technical details that one would consider in a completely rigorous derivation of a quantum filter.  In particular, we do not demonstrate that our result is optimal, nor even delve into the details of what optimal means in this context.  Suffice it to say, the results hold up under more rigorous mathematical scrutiny, and the interested reader is encouraged to consult Refs.\ \cite{vanHandel:2005a,Bouten:2006a} for a more thorough discussion of these details.  Readers who are familiar with mathematical filtering theory will miss little by only skimming the derivation as it is presented here.

The task of the quantum filter is to construct a best-estimate of atomic observables conditioned on continuous observation of the transmitted probe field.  Our first step is one of convenience---we rewrite the double-pass propagator of Eq.\ (\ref{eq:double_pass_prop}) in the general form \cite{Bouten:2006a}
\begin{equation} \label{eq:app:propagator}
    d\hat{U}_t = \left[ \hat{L} d\St^{\dag} - \hat{L}^{\dag} d\St
         - \frac{1}{2}\hat{L}^{\dag}\hat{L} dt -i\hat{H}_c dt\right]\hat{U}_t ,
\end{equation}  
identifying $\hat{L} = \sqrt{M}\Fz + i\sqrt{K}\Fy$ and $H_c = \hat{H} - \sqrt{KM}(\Fz\Fy + \Fy\Fz)/2 $.  Then, the Heisenberg-picture evolution of any atomic observable $\hat{X}$ is given by the so-called quantum flow
\begin{equation}
	j_t(\hat{X}) = \hat{U}_t^{\dag}(\hat{X} )\hat{U}_t.
\end{equation}

Applying the It\^{o} product rule (discussed in Appendix \ref{app:product_rule}) twice and noting the Schr\"{o}dinger picture operator satisfies $d\hat{X} = 0$, we find
\begin{eqnarray}
    dj_t(\hat{X}) &= & \hat{U}_t^{\dag}\hat{X} d\hat{U}_t 
                     + d\hat{U}_t^{\dag}\hat{X} \hat{U}_t
                     + d\hat{U}_t^{\dag}\hat{X} d\hat{U}_t  \label{eq:app:dynamics} \\
                    &=&  j_t([\hat{X},\hat{L}])d\St^{\dag} 
                      + j_t([\hat{L}^{\dag},\hat{X}])d\St
                      + j_t(\mathcal{L}(\hat{X}))dt, \nonumber
\end{eqnarray}
where we have defined the familiar Lindblad generator
\begin{equation}
    \mathcal{L}(\hat{X}) = i[\hat{H},\hat{X}] + \hat{L}^{\dag}\hat{X}\hat{L}
                         - \frac{1}{2}(\hat{L}^{\dag}\hat{L}\hat{X}
                                      +\hat{X}\hat{L}^{\dag}\hat{L}) .
\end{equation}
Similarly, the observation process corresponding to polarimetric detection of the $y$-component of the vector Stokes operator is given by
\begin{equation}
	Z_t = \hat{U}_t^{\dag}(\St^{\dag} + \St)\hat{U}_t .
\end{equation}  
Two applications of the It\^{o} product rule, which generate seven initial It\^{o} products, gives its time evolution as
\begin{equation} 
    \label{eq:app:observations}
    dZ_t = j_t(\hat{L} + \hat{L}^{\dag})dt 
               + d\St^{\dag} + d\St .
\end{equation}

In the language of classical control theory, Eqs.\ (\ref{eq:app:dynamics}) and (\ref{eq:app:observations}) are a system/observation pair.  We see that the observations process carries information about the state of the system, $j_t(\hat{L} + \hat{L}^{\dag})$, albeit corrupted by the quantum white noise term $d\St^{\dag} + d\St$.  Moreover, as seen in Eq.\ (\ref{eq:app:dynamics}), the system itself is also driven by the quantum white noises.  The quantum filter is tasked with picking out the relevant information from the observations and combining it with the model of the dynamics in order to estimate system observables.

Mathematically, the filter is given by the conditional expectation 
\begin{equation} \label{Equation::FilterCE}
    \pi_t[\hat{X}] = \mathbbm{E}[j_t(\hat{X}) | Z_{(0,t)}],
\end{equation}
where $Z_{(0,t)}$ is the entire measurement record up to time $t$.  Intuitively, we expect that the filter will be some function of the measurement process on which we are conditioning.  This statement can be made more mathematically rigorous by showing that the conditional expectation in Eq.\ (\ref{Equation::FilterCE}) is adapted to the measurement process.  Furthermore, it is possible to derive a recursive form for the filter, in which only the most recent measurement increment is needed to update our estimate, and the dynamical equation for the filter can be written as
\begin{equation}
    d\pi_t[\hat{X}] = k_t(\hat{X})dt + m_t(\hat{X}) dZ_t ,
\end{equation}
where $k_t(\hat{X})$ and $m_t(\hat{X})$ are yet to be determined functions and the form is the most general we could consider, given that we only have access to the measurements.  In order to find the forms of these functions, we can utilize the property of conditional expectation
\begin{equation}  \label{eqn:cet}
    \mathbbm{E}[\hat{Y}\mathbbm{E}[\hat{X}|Z_{(0,t)}]]
    = \mathbbm{E}[\hat{Y}\hat{X}] ,
\end{equation}
which is true for any $\hat{Y}$ that is a function of the measurement process.  

One way of enforcing Eq.\ (\ref{eqn:cet}) is by ensuring that the following holds:
\begin{equation} \label{eq:app:filter_equality}
	\mathbbm{E}\left[\pi_t[\hat{X}] e^{\int_0^t g(s)dZ_s}\right] = \mathbbm{E}\left[j_t(\hat{X})e^{\int_0^t g(s)dZ_s}\right],
\end{equation}
since by taking linear combinations and appropriate derivatives of either side, any (analytic) function $\hat{Y}$ of the measurement current can be generated via an appropriate choice of $g(t)$.  To simplify the algebraic manipulations, we multiply both sides by the deterministic integrating factor $\exp(-\frac{1}{2}\int_0^t g(s)^2 ds)$ and identify our ``generating function'' as
\begin{equation}
	e^g_t  = \exp\left(\int_0^t g(s)dZ_s - \frac{1}{2} \int_0^t g(s)^2 ds\right) .
\end{equation}
To calculate the stochastic differential equation for $e_t^g$, we find it useful to set $d\hat{R}_t = g(t)dZ_t - \frac{1}{2}g(t)^2dt$, so that $e_t^g = \exp(\hat{R}_t)$.  Using the chain rule and expanding to second order in differentials, we have
\begin{eqnarray}
    de_t^g &=&  e_t^g d\hat{R}_t + \frac{1}{2} e_t^g (d\hat{R}_t)^2\\
           &=& g(t)e_t^g dZ_t  ,   
\end{eqnarray}
where we have used the fact that $(d\hat{R}_t)^2 = (g(t)dZ_t)^2 = g(t)^2d\St d\St^{\dag} = g(t)^2dt$.  

Since the individual white noise terms, $d\hat{S}_t,d\hat{S}_t^{\dag}$, are zero in expectation, we can formally take the time derivative of either side of Eq.\ (\ref{eq:app:filter_equality}) as long as we use the It\^{o} product rule of Eq.\ (\ref{eq:app:ito_product}) prior to taking the expectation.  Doing so, we find
\begin{eqnarray}
    \frac{d}{dt}\mathbbm{E}[e_t^g\pi_t(\hat{X})] &=&  \mathbbm{E}\left[e_t^g\left(k_t(\hat{X}) + j_t(L + L^{\dag})m_t(\hat{X})\right)\right. \label{eq:app:pitderiv} \\
    && \left. + g(t)e_t^g\left(m_t(\hat{X}) + j_t(L+L^{\dag})\pi_t[\hat{X}]\right)\right] \nonumber \\
    \frac{d}{dt}\mathbbm{E}[e_t^gj_t(\hat{X})] &=&
  	 \mathbbm{E}\left[ e_t^g j_t(\mathcal{L}[\hat{X}]) \right. \label{eq:app:jtderiv} \\
	 && \left. 
  	      \quad\quad + g(t)e_t^gj_t(\hat{L}^{\dag}\hat{X} + \hat{X}\hat{L})\right] \nonumber
\end{eqnarray}
Note that in reaching the above form of Eq.\ (\ref{eq:app:jtderiv}), we made use of the identities
\begin{align}
   j_t(\hat{A})j_t(\hat{B}) &= j_t(\hat{A}\hat{B}) \\
   j_t(\hat{A}) - j_t(\hat{B}) &= j_t(\hat{A} - \hat{B}), 
\end{align}
which are readily verified using the definition of $j_t(\hat{X})$.

To obtain the quantum filtering equation, we equate like terms in Eqs.\ (\ref{eq:app:pitderiv}) and (\ref{eq:app:jtderiv}) to enforce Eq.\ (\ref{eq:app:filter_equality}), and to ensure that $m_t$ and $k_t$ remain functions of the measurement record, we replace $\mathbbm{E}[j_t(\cdot)]$ with $\mathbbm{E}[\pi_t(\cdot)]$, as this is just Eq.\ (\ref{eq:app:filter_equality}) with $g(t) = 0$.  After a little rearranging, we  arrive at the \emph{quantum filtering equation}
\begin{eqnarray} \label{eq:app:quantum_filter_eq}
    d\pi_t[\hat{X}]  & = & \pi_t[\mathcal{L}[\hat{X}]]dt  \\
    & & + \left(\pi_t[\hat{L}^{\dag}\hat{X} + \hat{X}\hat{L}]
    - \pi_t[\hat{L}^{\dag} + \hat{L}]\pi_t[\hat{X}]\right) \nonumber \\
 & &   \times \left(dZ_t - \pi_t[\hat{L} + \hat{L}^{\dag}] dt\right) \nonumber
\end{eqnarray}

Oftentimes, and particularly in the traditional quantum optics setting, one works with the so-called adjoint form for the filter, which gives a dynamical equation for a density matrix $\rho_t$ that satisfies $\pi_t[\hat{X}] = \Tr{\hat{X}\rho_t}$.  This is easily determined from Eq.\ (\ref{eq:app:quantum_filter_eq}).  Plugging in the specific $\hat{L}$ and $\hat{H}$ for our double-pass system undergoing Larmor precession, we find the \emph{double-pass quantum filter}
\begin{eqnarray} 
    d\rho_t & = & i\gamma B[\Fy,\rho_t]dt + i\sqrt{KM}[\Fy,\{\Fz,\rho_t\}]dt \nonumber \\
	&&    + M \mathcal{D}[\Fz]\rho_t dt 
             + K\mathcal{D}[\Fy]\rho_t dt \label{eq:adjoint_quantum_filter} \\
         &&
             + \left(\sqrt{M}\mathcal{M}[\Fz]\rho_t 
             + i\sqrt{K}[\Fy,\rho_t]\right)dW_t \nonumber
\end{eqnarray}
where the innovations process 
\begin{equation}
	dW_t = dZ_t - 2\sqrt{M}\Tr{\Fz\rho_t}dt
\end{equation}
is a Wiener process, i.e. $\mathbbm{E}[dW_t] = 0, dW_t^2 = dt$.  The various superoperators are defined as
\begin{align}
    \mathcal{D}[\Fk]\rho_t &= \Fk\rho_t\Fk^{\dag} - \frac{1}{2}\Fk^{\dag}\Fk\rho_t - \frac{1}{2}\rho_t\Fk^{\dag}\Fk\\
    \mathcal{M}[\Fz]\rho_t &= \Fz\rho_t + \rho_t\Fz - 2\Tr{\Fz\rho_t}\rho_t\\
    \{\Fz,\rho_t\} &= \Fz\rho_t + \rho_t\Fz
\end{align}
In looking at the structure of the filter, we see a deterministic term which evolves the current state estimate via the standard open system evolution one would expect for the system.  The stochastic term performs the conditioning and is weighted by the innovations process, which captures how much new information the latest measurements provide.  

One other form which is useful when the quantum state remains pure is the stochastic Schr\"{odinger} equation (SSE).  As developed in Appendix \ref{app:SSE}, the SSE for the double-pass quantum filter is
\begin{eqnarray} \label{eq:double_pass_SSE}
	d\ket{\psi}_t & = & \left(
		i\gamma B\Fy -\frac{M}{2}(\Fz-\expect{\Fz}_t)^2 \right. \\
	&& \left. + i\sqrt{KM}\Fy(\Fz + \expect{\Fz}_t)
	 					- \frac{K}{2}\Fy^2\right)\ket{\psi}_t dt \nonumber \\
	&&
				  + \left( \sqrt{M}(\Fz - \expect{\Fz}_t) + i\sqrt{K}\Fy\right)\ket{\psi}_t dW_t . \nonumber
\end{eqnarray}

\section{The Quantum Cram\'{e}r-Rao Inequality } \label{section:cr}
In order to characterize the performance of the magnetometer, we consider quantum information theoretic bounds on the units-corrected mean-square deviation of the magnetic field estimate $\tilde{B}$ of the true magnetic field $B$ \cite{Braunstein:1994a,Braunstein:1995a},  given in Eq.\ (\ref{Equation::CRError}).  The quantum Cram\'{e}r-Rao bound  \cite{Helstrom:1976a,Holevo:1982a,Braunstein:1994a,Braunstein:1995a} states that the deviation of \emph{any} estimator is constrained by  
\begin{equation}
    \delta \tilde{B} \geq \frac{1}{\sqrt{\mathcal{I}_B(t)}}, \quad \mathcal{I}_B(t) = \Tr{\rho_B(t)\mathfrak{L}_B^2(t)} ,
\end{equation}
where the quantum Fisher information $\mathcal{I}_B(t)$ is the expectation of the square of the symmetric logarithmic derivative operator, defined implicitly as
\begin{equation}
    \frac{\partial \rho_B(t)}{\partial B} = \frac{1}{2}(\mathfrak{L}_B(t)\rho_B(t) + \rho_B(t)\mathfrak{L}_B(t)) .
\end{equation}
For pure states, $\rho_B^2 = \rho_B$, so that
\begin{equation}
    \mathfrak{L}_B(t) = 2\frac{\partial \rho_B(t)}{\partial B}
\end{equation}
which indicates
\begin{equation} \label{eq:cramer_rao}
    \delta \tilde{B} \geq \frac{1}{2} \left\langle\left(\frac{\partial \rho_B(t)}{\partial B}\right)^2\right\rangle^{-\frac{1}{2}} .
\end{equation}
In this form, we see that the lower bound is related to the sensitivity of the evolved state to the magnetic field parameter.  That is, any estimator's performance is constrained by how well the dynamics transform differences in the value of $B$ into differences in Hilbert space.  

As discussed by Boixo et.~al in \cite{Boixo:2007a}, for Hamiltonian evolution, the quantum Cram\'{e}r-Rao bound may be expressed in terms of the operator semi-norm, which is the difference between the largest and smallest (non-degenerate) eigenvalues of the probe Hamiltonian.  For the magnetic dipole Hamiltonian in Eq.\ (\ref{Equation::HLarmor}), this bound is simply the Heisenberg limit in Eq.\ (\ref{Equation::DeltaBHL}).  More generally, the authors show that a probe Hamiltonian which involves $k$-body operators gives rise to an uncertainty scaling of $1/tF^k$.  They further argue that no ancillary quantum systems or auxiliary Hamiltonians contribute to this bound; it is determined solely by the Hamiltonian that directly involves the parameter of interest.

Such analysis suggests the double-pass quantum system, whose only direct magnetic field coupling is in the magnetic dipole Hamiltonian, should show no more sensitivity than a single pass system.  There are several reasons why we believe there is more to the story.  Firstly, the unitary evolution of the joint atom-field system in Eq.\ (\ref{eq:app:propagator}) involves an auxiliary system of infinite dimension.  As such, it is not clear that the arguments leading to the operator semi-norm are valid, in particular due to the fact that the white noise terms $d\St,d\St^{\dag}$ are singular.  Additionally, the double-pass limit is a Markov one, in which the interaction the light field mediates between atoms is essentially instantaneous relative to other time-scales in the problem.  The effective interaction is therefore fundamentally different than one in which measurements of a finite dimensional ancilla system are used to modulate the evolution of the probe atoms.  Thus the conditioned system, given in terms of the quantum filter of Eq.\ (\ref{eq:app:quantum_filter_eq}) or (\ref{eq:double_pass_SSE}), does not correspond to unitary dynamics.  Indeed, looking at Eq.\ (\ref{eq:double_pass_SSE}), we see that the local generator of dynamics is \emph{path-dependent}, given in terms of the expectation of $\Fz$.  Therefore, as the magnetic field directly impacts the state through the magnetic dipole term, it also non-trivially modulates future dynamics through a state-dependent generator.

\subsection{Numerical Analysis of the Quantum Fisher Information}

\begin{figure*}[t]
    \centering
        \includegraphics[scale=1]{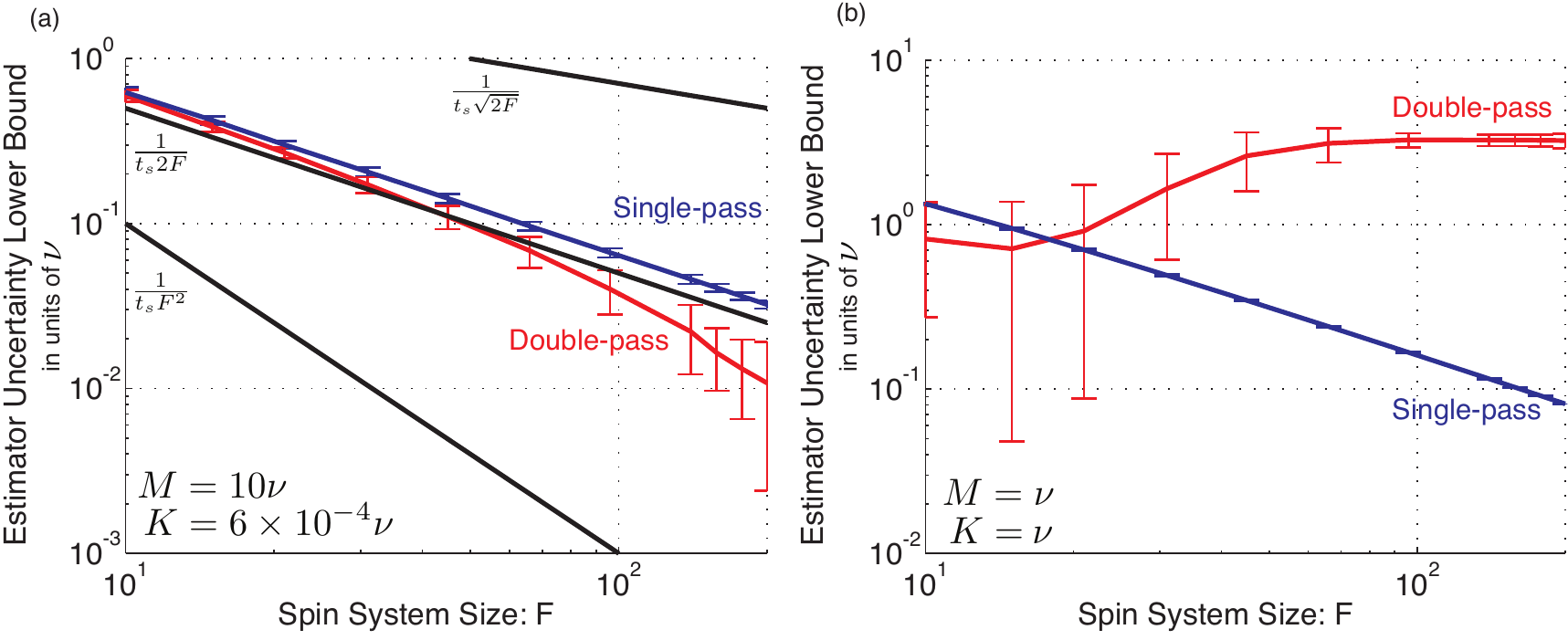}
    \caption{(color online) Comparison of the lower bound on estimator uncertainty for the single- and double-pass systems determined by finite-difference calculations of the quantum Fisher information.  Both plots approximate the derivative in Eq.\ (\ref{eq:cramer_rao}) using $\delta B = 5\times 10^{-4}\nu$ at a final integration time $t_s = 0.1\nu^{-1}$.  The lower bound estimate is averaged over 100 simulated trajectories with the error bars indicating the standard deviation observed in those 100 samples. (a) Lower bound scaling for $M = 10\nu, K = 6\times 10^{-4}\nu$.  A least-squares fit shows that the single-pass uncertainty is consistent with the Heisenberg scaling, with a numerical power law fit of $F^{-0.99}$.  The double-pass clearly improves faster than $1/F$ over the range of $F$ considered here.   Also shown for comparison are the analytic shot-noise, Heisenberg and two-body coupling lower bounds. (b) Simulations were also performed for a difference choice of double-pass coupling parameters $M = K = 1\nu$,  where it can be seen that the improved scaling is highly dependent on selecting appropriate coupling strengths. }
    \label{fig:cramer_rao}
\end{figure*}

Unfortunately, it is not clear how to fold the quantum stochastic or quantum filtered dynamics analytically into the semi-norm bound considered in \cite{Boixo:2007a}.  Nonetheless, the quantum Cram\'{e}r-Rao bound in Eq.\ (\ref{eq:cramer_rao}) is excellent fodder for computer simulation.  By numerically integrating the stochastic Schr\"{o}dinger form of the quantum filter in Eq.\ (\ref{eq:double_pass_SSE}), a finite difference approximation of $\partial \rho_B(t)/\partial B$ may be evaluated for different collective spin sizes $F$.  That is, for a given choice of $F$, a finite difference approximation of the quantum Fisher information near $B = 0$ can be constructed by co-evolving three trajectories, $\rho_0(t_s)$, $\rho_{\delta B}(t_s)$, and $\rho_{-\delta B}(t_s)$ (seeded by the same noise realization),  and calculating
\begin{equation}
    \left< \left(\frac{\partial \rho_B(t_s)}{\partial B}\right)^2 \right>
        \approx \Tr{\left(\frac{\rho_{\delta B}(t_s) 
                - \rho_{-\delta B}(t_s)}{2\delta B}\right)^2 \rho_{0}(t_s)}.
                \label{eqn:cr_fd}
\end{equation}   
The expectation value is then evaluated numerically by averaging over a statistical ensemble of such measurement \textit{trajectories.}

\subsubsection{Simulation Results}

Figure \ref{fig:cramer_rao} shows the results of such simulations of the quantum Fisher information as a function of the spin size $F$ and different relative values of the coupling strengths $M$ and $K$.  For convenience, all rates are taken to be unitless, defined in units of an arbitrary frequency $\nu$ (which is then set to $\nu = 1$ in the actual simulation).  For all cases described here, the quantum filter in Eq.\ (\ref{eq:double_pass_SSE}) was integrated from an initially $x$-polarized spin coherent state using a second-order predictor-corrector stochastic integrator \cite{Kloeden:1992a} with a step size $dt = 10^{-5}\nu^{-1}$ until a time $t_s = 0.1\nu^{-1}$.   The density operators  in Eq.\ (\ref{eqn:cr_fd}) were constructed as $\rho_B(t_s) = \ketbra{\psi_B(t_s)}{\psi_B(t_s)}$ for the state of the system at time $t_s$ evolved under the magnetic field with value $B$.  The finite-difference parameter was chosen to be $\delta B = 5\times 10^{-4}\nu$ and $\gamma=1$.  Due to the stochastic nature of the evolution, the finite difference approximation was averaged over 100 trajectories, with each member of the three density matrices coevolved using the same white noise realizations.  The plots in Fig.~\ref{fig:cramer_rao} correspond to the lower bound determined by Eq.\ (\ref{eq:cramer_rao}) using the finite difference approximation and finite averaging approximation just described.  The error bars correspond to the standard deviation of the lower bound for the ensemble of 100 trajectories.  A similar finite difference approximation around the mean value $B = 0.1\nu$ was conducted and showed the same behavior as that described here for a mean value of $B=0$.

Figure \ref{fig:cramer_rao}(a) plots the estimator uncertainty lower bound as a function of $F$ for a double-pass system, with $M = 10\nu$ and $K = 6\times 10^{-4}\nu$, and an equivalent single-pass system with $M=10\nu$ and $K = 0$.  For these parameter values, the simulations show a clear improvement in the scaling of the uncertainty lower bound, even taking into account the increased fluctuations of the double-pass system.  Also shown are the analytic scalings for shot-noise uncertainty Eq.\ (\ref{Equation::DeltaBSN}), Heisenberg uncertainty Eq.\ (\ref{Equation::DeltaBHL}) (with $\alpha = 1 $) and the quantum Cram\'{e}r-Rao bound for a hypothetical magnetic field coupling interaction $-\hbar\gamma B \Fy^2$. A least-squares fit to a power-law form suggest the single-pass system scales as $F^{-0.99}$, which is in good agreement with the Heisenburg-limited scaling of $F^{-1}$.  A power law fit for the double-pass system gives an exponent of $-1.34$, although its lower bound appears to have more complicated scaling.    In fact, the behavior towards larger $F$ suggests that the double-pass system may show a super-polynomial uncertainty scaling. 

We note that there is a constant prefactor difference between our estimator simulations and the conventional Heisenberg limit.  This prefactor appears in the plot as a constant offset between the simulated single-pass bound and the $1/t_s2F$ Heisenberg bound.  There are two reasons why one might expect such an offset:  (1) the Heisenberg limit is computed from the Cram\'{e}r-Rao inequality by optimizing over all initial states, including optimally squeezed states or ``cat'' superposition states which achieve the bound while our procedure begins from a spin coherent state which is then gradually entangled by the measurement process; (2) a continuous measurement scenario based on the theory of open quantum systems cannot be expressed in the form of a simple Hamiltonian.

Furthermore, our simulations reveal that the uncertainty scaling of the initially $x$-polarized spin coherent state depends sensitively on the particular choice of coupling strengths $M$ and $K$, as seen in Fig.\ \ref{fig:cramer_rao}(b).  This figure considers the case of equal coupling strengths, $M = K = \nu$, and we see that the double-pass system quickly loses its advantage after $F\approx 25$.  This can be understood in light of the quantum stochastic model of the previous subsection.  In considering the general stochastic propagator of Eq.\ (\ref{eq:app:propagator}), we identified the coupling operator $\hat{L} = \sqrt{M}\Fz + i\sqrt{K}\Fy$, which if $M = K$, is essentially the angular momentum lowering operator along $x$---$\Fmx$.  If $M,K \gg \gamma B$, a continuous measurement of this operator very quickly moves the $+x$-polarized initial state onto the $-x$-polarized state, which is an attractive fixed point of $\Fmx$.  Once this state is reached, the dynamics become relatively insensitive to the magnetic field value and result in a poor uncertainty lower bound.  On the other hand, if $M,K$ are much smaller than $\gamma B$, the positive feedback from the $i\sqrt{K}\Fy$ term is washed out by Larmor precession.  Given that we are interested in detection limits, i.e. $B \approx 0$, we do not focus on the regime where Larmor precession dominates.

Regardless, Figure \ref{fig:cramer_rao}(b) suggests that for a given $M$ and $K \neq 0$, there will be some value of $F$ after which the lowering operator dynamics dominate, rendering the double-pass system relatively useless for estimating $B$.  But by manipulating $M$ and $K$ relative to $F$, the extra dynamics allow for an improved sensitivity to the magnetic field.  Indeed, Fig.~\ref{fig:cramer_rao}(a) uses a very small second-pass strength $K$, relative to the first-pass strength $M$, so that just the right amount of positive feedback enters via the $i\sqrt{K}\Fy$ term in the measured operator.  The positive feedback can be viewed as increasing the magnetic fields effect on the rate of precession.

In short, Figure \ref{fig:cramer_rao}(a) suggests that there are some parameter values, appropriate for some range of $F$, which show an estimator uncertainty lower bound scaling below the Heisenberg limit.  It is important to note that computational constraints limited simulations to $F \approx 150$, which is well below the collective spin size one might expect for $10^4-10^6$ atoms.  However, we believe the scalings are nonetheless suggestive of quite general behavior for this system.  Although in practice, it seems that one would need to fine tune the coupling strengths $M$ and $K$ in order to be in a regime with such scaling.  It may be that such coupling strengths are inaccessible in an experimental setting.  While this is an important consideration, there is a more pressing theoretical question---does a practical estimator exist which saturates the quantum Cram\'{e}r-Rao bound?  We summarize our search for such an estimator in the following section.

\section{Magnetic Field Estimators} 
\label{sec:magnetic_field_estimators}

While studying the properties of lower bounds on estimator performance is important for developing an understanding of the capabilities of a given parameter coupling scheme, any actual procedure for implementing quantum parameter estimation must also develop a constructive procedure for doing the estimation.

In this section, we consider two methods for estimating the strength of the magnetic field $B$ based on the stochastic measurement record $Z_{(0,t)}$.  In both cases, we extend the quantum filters developed in the previous section to account for our uncertainty in $B$, which in turn results in new filters capable of estimating $B$.  We note that our approach is similar in spirit to Refs.~\cite{Molmer:2004a,Geremia:2003a}. 

\subsection{Quantum Particle Filter} 
\label{sec:quantum_particle_filter}
The technique of quantum particle filtering, as developed in Ref.~\cite{Chase:2008a}, leverages the fact that the quantum filtering equations already provide a means for estimating the \emph{state} of a quantum system conditioned on the measurement record.  If we place the magnetic field parameter on the same footing as the quantum state, we can simply apply the quantum filtering results we already derived.  Indeed, by embedding the magnetic field parameter as a diagonal operator in an auxiliary Hilbert space, the quantum filter \emph{still} gives the best estimate of \emph{both} system and auxiliary space operators.  We accomplish this by promoting the magnetic field parameter to the diagonal operator
\begin{equation}
    B \mapsto \hat{B} = \int B \ketbra{B}{B} dB \in \mathcal{H}_B, 
\end{equation}
where $\mathcal{H}_B$ is the new auxiliary Hilbert space with basis states satisfying $\hat{B}\ket{B} = B\ket{B}$ and $\braket{B}{B'} = \delta(B-B')$.  All atomic operators and states, which are associated with the atomic Hilbert space $\mathcal{H}_A$, act as the identity on $\mathcal{H}_B$, e.g. $\Fz \mapsto I \otimes \Fz$.  The only operator which joins the two spaces is the magnetic dipole Hamiltonian, which is now given by
\begin{equation}
    \hat{H} \mapsto -\hbar\gamma \hat{B}\otimes\Fy
\end{equation}
The derivation of the quantum filtering equation is essentially unchanged, provided one replaces atomic operators with the appropriate forms for the joint space $\mathcal{H}_B \otimes \mathcal{H}_A$.  

For parameter estimation, the adjoint form is the more convenient version of the quantum filter.  Since $\hat{B}$ corresponds to a classical parameter, we require the marginal density matrix $(\rho_B)_t = \operatorname{Tr}_\mathcal{H_A}[\rho_t]$ be diagonal in the basis of $\hat{B}$, so that it corresponds to a classical probability distribution.  This suggests we write the total conditional density matrix in the ensemble form
\begin{equation} \label{eq:ensemble_continuous_form}
    \rho_t^E = \int dB p_t(B) \ketbra{B}{B} \otimes \rho_t^{(B)} 
\end{equation}  
where $p_t(B) = P(B | Z_{(0,t)})$ is precisely the conditional probability density for $B$.

While one could attempt to update this state via the quantum filter, doing so is entirely impractical, as one can not represent an arbitrary distribution for $p_t(B)$ with finite resources.  Instead, one approximates the distribution with a weighted set of point masses or \emph{particles}:
\begin{equation} \label{eq:approximate_density}
    p_t(B) \approx \sum_{i=1}^N p_t^{(i)} \delta(B - B_i) .
\end{equation}
The approximation can be made arbitrarily accurate in the limit of $N \rightarrow \infty$.  Plugging this distribution into the ensemble density matrix form of Eq. \ref{eq:ensemble_continuous_form} gives
\begin{equation} \label{eq:discrete_ensemble_rho}
    \rho_t^E = \sum_{i = 1}^N p_t^{(i)} \ketbra{B_i}{B_i} \otimes \rho_t^{(B_i)} 
\end{equation}
Each of the $N$ triples $\{p_t^{(i)}, B_i, \rho_t^{(B_i)} \}$ is called a \emph{quantum particle}.  Intuitively, the particle filter works by discretizing the parameter space and then evolving an ensemble of quantum systems according to the exact dynamics for each parameter value.  The filtering equations below perform Bayesian inference on this ensemble, updating the relative probabilities of particular parameter values given the measurement record.

The quantum particle filter for the double-pass system with \emph{unknown} $B$ is found by plugging the discretized ensemble $\rho_t^E$ into the extended double-pass filter.  After a little manipulation, one finds
\begin{subequations}\label{eq:ensemble_filtering_eqs}
	\begin{eqnarray} 
		    dp_t^{(i)} &= & 2\sqrt{M}(\Tr{\Fz\rho_t^{(B_i)}} \nonumber \\
		    &&
		        - \sum_{j=1}^N p_t^{(j)}\Tr{\Fz\rho_t^{(B_j)}})p_t^{(i)}dW_t 
		          \\
		    d\rho_t^{(B_i)} &= & i\gamma B_i[\Fy,\rho_t^{(B_i)}]dt
		     + \sqrt{KM}[\Fy,\{\Fz,\rho_t^{(B_i)}\}]dt \nonumber\\
		    && + M \mathcal{D}[\Fz]\rho_t^{(B_i)} dt 
		             + K\mathcal{D}[\Fz]\rho_t^{(B_i)} dt \\
		             && + \left(\sqrt{M}\mathcal{M}[\Fz]\rho_t^{(B_i)} 
		             + i\sqrt{K}[\Fy,\rho_t^{(B_i)}]\right)dW_t \nonumber \\
		    dW_t &=& dZ_t 
		        - 2\sqrt{M}\sum_{i=1}^Np_t^{(i)}  \Tr{\Fz\rho_t^{(B_i)}}dt
	\end{eqnarray}
\end{subequations}
where the prior distribution $p_0(B)$ is used to determine the initial parameter weights, $p_0^{(i)}$, and values, $B_i$.  All initial quantum states, $\rho_0^{(B_i)}$, are taken to be the spin coherent state pointing along $+x$.

Looking at the structure of this filter, we see that each particle's quantum state $\rho_t^{(B_i)}$ evolves under the standard quantum filter we would use if we knew $B = B_i$.  The innovations process $dW_t$ serves to couple the different particles, since it depends on the ensemble average of the measurement observable.

An estimate of the magnetic field strength is then constructed from the approximate density in Eq.\ (\ref{eq:approximate_density}), either taking the most probable $B$ value, corresponding to the largest $p_t^{(i)}$ or calculating the expected value of $\hat{B}$
\begin{equation}
    \tilde{B}_{pf} = \expect{\hat{B}} = \sum_{i=1}^{N} p_t^{(i)} B_i .
\end{equation}
For the latter estimate, the uncertainty is given by
\begin{eqnarray}
    \Delta \tilde{B}_{pf} & = & \left(\expect{\hat{B}^2} - \tilde{B}_{pf}^2\right)^{1/2}  \nonumber \\
    & = & \left(\sum_{i=1}^{N} p_t^{(i)} B_i^2 - \tilde{B}_{pf}^2\right)^{1/2}  .
\end{eqnarray}

One important feature of particle filters which we will mention, but not utilize in this paper, is the potential for resampling.  As presented, the initial particle ensemble $\{p_0^{(i)}, B_i, \rho_0^{(B_i)} \}$ fixes the possible values of $B_i$ at the outset.  If there are no initial points near the true value of $B$, the filter will by construction have difficulty finding this value.  Even if one of the particles is associated with the true value of $B$, the filter still integrates the dynamics for a potentially large number of low weight particles which contribute little to the estimate.  A way around this issue is to resample and create new particles during the course of integration, either drawing from the discrete distribution $\{p_t^{(i)}\}$ or using finer and finer regular grids to hone in on important regions of parameter space.  But in both cases, it is not clear how to resample the associated quantum states $\rho_t^{(B_i)}$ in order to improve the estimate.  Rather than considering such options here, we refer the reader to \cite{Chase:2008a,Doucet:2001} for more discussion.

Finally, we note that the computational resources required for particle filtering are demanding, generally requiring space and time which scale as $N{(\dim{\mathcal{H}_A}})^2$, though we can get down to $N{(\dim{\mathcal{H}_A}})$ if the atomic state remains pure.  Moreover, the particle filter is biased for any finite number of particles, although the variance of estimates of random variables using $\{p_t^{(i)}\}$ converges as $N^{-1}$ \cite{Doucet:2001}.  Although this will not prohibit us from exploring the magnetic field sensitivity scaling of systems with collective spins up to $F = 140$, using a large particle set is entirely impractical for real systems with $F$ orders of magnitude larger.  For such systems, we need a low-dimensional state representation whose size does not scale with $F$, after which the particle filter becomes a more promising technique for parameter estimation.  We explore low dimensional representations of the double-pass system in the following subsection.

\subsection{Quantum Kalman Filter} 
\label{sec:quantum_kalman_filter}
Rather than constructing a magnetic field estimator from the exact quantum dynamics, one could instead first focus on deriving an approximate filter for the atomic state, which is then a starting point for the magnetic field estimator.  Indeed, previous work in precision magnetometry via continuous measurement \cite{Geremia:2003a} has taken this route by constructing a \emph{quantum Kalman filter} to describe the atomic dynamics.  Such a filter leverages the fact that for an initially spin polarized state of many atoms (say along $+x$), a first order Holstein-Primakoff expansion \cite{Holstein:1940a} linearizes the small-angle dynamics in terms of a Gaussian state characterized by the means $\pi_t[\Fz], \pi_t[\Fy]$ and the covariances $\Delta\Fz^2,\Delta\Fy^2,\Delta\Fz\Fy$.  Just as in classical filtering theory, the conditional state for a linear system with Gaussian noise is itself described by a Gaussian distribution and therefore only requires filtering equations for the means and a deterministic equation for the variances \cite{Kalman:1960a,Kalman:1961a}.  For the case of magnetometry, the number of these parameters is independent of the number of atoms in the atomic ensemble.  We will also find that within this approximation, we can again embed $B$ as an unknown state parameter and find a corresponding Kalman filter appropriate for estimating its value.  

However, applying the small-angle and Gaussian approximations in the quantum case is usually done in an ad-hoc fashion, especially in light of the recent introduction of \emph{projection filtering} into the quantum filtering setting \cite{vanHandel:2005b,Mabuchi:2008a}.  In this framework, one selects a convenient manifold of states whose parameterization reflects the approximations to enforce.  At each point in this manifold, the exact differential dynamics induced on these states is orthogonally projected back into the chosen family.  For our purposes, this means projecting the filter in Eq.\ (\ref{eq:double_pass_SSE}) onto a manifold of Gaussian spin states. Although the resulting equations are not substantively different than those derived less carefully, we believe the potential application of projection filtering in deriving other approximate filters and master equations warrants the following exposition.  

\subsubsection{Projection Filter Overview}
Abstractly, projection filtering proceeds as follows.  We assume we already have a dynamical equation, such as Eq.\ (\ref{eq:double_pass_SSE}), for a given manifold of states, such as pure states.  For convenience, let these dynamics be represented as
\begin{equation}
    d\ket{\psi}_t = \mathcal{N}[\ket{\psi}_t] ,
\end{equation}
where $\mathcal{N}$ is the generator of dynamics.  Now select the desired family of ``approximating'' states which are a submanifold of the exact states.  We assume this family is parameterized by a finite number of parameters $x_1,x_2,\ldots,x_n$ and we denote states in this family as $\ket{x_1,x_2,\ldots,x_n}$.  At every point in this manifold, the tangent space is spanned by the tangent vectors
\begin{equation}
    v_i = \partialD{\ket{x_1,x_2,\ldots, x_n}}{x_i} .
\end{equation}
Loosely speaking, these tangent vectors tell us how differential changes in the parameters move us through the corresponding submanifold of $\ket{x_1,x_2,\ldots,x_n}$ states in the space of pure states.  This is particularly useful, as the action of the generator $\mathcal{N}[\ket{x_1,x_2,\ldots,x_n}]$ does not necessarily result in a state within the family.  But by projecting the dynamics onto the tangent space, we can find a filter, called the projection filter, which constrains evolution within the chosen submanifold.  Explicitly, this projection is written as
\begin{eqnarray} 
    T & = & \Pi_{\text{span}\{v_i\}}[d\ket{x_1,x_2,\ldots,x_n}] \nonumber \\
        & = & \sum_i \frac{\langle v_i,
            \mathcal{N}[\ket{x_1,x_2,\ldots,x_n}] \rangle}
                {\langle v_i, v_i \rangle}v_i, \label{eq:general_projection}
\end{eqnarray}
where in this pure state formulation, the inner product is the standard Hilbert space inner product.  

\subsubsection{Gaussian State Family and Tangent Vectors}
For our double-pass magnetometer, we begin by introducing the two-parameter family of Gaussian states
\begin{align} \label{eq:gaussian_state_parameterization}
        \ket{\theta_t,\xi_t} &= e^{-i\theta_t\Fy}e^{-2i\xi_t(\Fz\Fy + \Fy\Fz)}\ket{F,+F_x} \nonumber\\
        &= \Yt\Sqt\ket{F,+F_x}, 
\end{align}
where $\ket{F,+F_x}$ is the spin coherent state pointing along $+x$, $\Sqt$ is a spin squeezing operator \cite{Kitagawa:1993a} with squeezing parameter $\xi_t$ and $\Yt$ is a rotation about the $y$-axis by angle $\theta_t$.  Intuitively, the squeezing along $z$ generated by $\Sqt$ corresponds to the squeezing induced by measuring $\Fz$.  The rotation via $\Yt$ then accounts for both the random evolution due to the measurement as well as any rotation induced by the magnetic field.  The tangent vectors for these states are
\begin{eqnarray}
    v_{\theta_t} & = & \partialD{\ket{\theta_t,\xi_t}}{\theta_t} \nonumber \\
    & = &   -i\Fy \Yt\Sqt\ket{F,+F_x}\\
    v_{\xi_t} & = & \partialD{\ket{\theta_t,\xi_t}}{\xi_t} \nonumber \\
    & = &
        \Yt\Sqt
        (-2i(\Fz\Fy + \Fy\Fz))\ket{F,+F_x} .
\end{eqnarray}

In calculating the normalization of these tangent vectors, we encounter terms such as
\begin{equation}
    \langle v_{\theta_t}, v_{\theta_t} \rangle
     = \bra{F,+F_x}\Sqt^{\dag}\Fy^2\Sqt\ket{F,+F_x}.
\end{equation}
More generally, almost all inner-products needed for the projection filter will be of the form
\begin{equation}
    \bra{F,+F_x} \Sqt^{\dag}g(\Fx,\Fy,\Fz)\Yt^{\dag} f(\Fx,\Fy,\Fz)
            \Yt\Sqt\ket{F,+F_x} . \nonumber
\end{equation}
Here, $g$ and $f$ are polynomial functions of their arguments.  Since $\Yt$ is a rotation, we can exactly evaluate
\begin{equation}
    \Yt^{\dag}f(\Fx,\Fy,\Fz) \Yt 
       = f( \Yt^{\dag}\Fx \Yt, 
           \Yt^{\dag}\Fy\Yt, \Yt^{\dag}\Fz\Yt), \nonumber
\end{equation}
where
\begin{eqnarray}
    \Yt^{\dag}\Fx\Yt &= & \Fx(\theta_t) = \Fx \cos{\theta_t} + \Fz\sin{\theta_t} \nonumber \\
    \Yt^{\dag}\Fy\Yt &= & \Fy\\
    \Yt^{\dag}\Fz\Yt &= & \Fz(\theta_t) = \Fz \cos{\theta_t} - \Fx\sin{\theta_t} . \nonumber
\end{eqnarray}
This leaves us with expectations of the form
\begin{equation} \label{eq:hp_expectations}
    \bra{F,+F_x}
    \Sqt^{\dag}g(\Fx,\Fy,\Fz)
            f(\Fx(\theta_t),\Fy,\Fz(\theta_t))
            \Sqt \ket{F,+F_x}
\end{equation}
where $g\times f$ will just be linear combinations of powers and products of $\Fx,\Fy,\Fz$.  Unfortunately, we cannot evaluate this expectation for arbitrary $\xi_t$.  However, for small $\xi_t$, the state which we are taking expectations with respect to is the ``squeezed vacuum'' in our preferred basis, e.g. it is the state $\ket{F,+F_x}$ pointing in the same direction, but with squeezed uncertainty in $\Fz$ and increased uncertainty in $\Fy$.  

For large $F$, angular momentum expectations of such a state are extremely well described by the Holstein-Primakoff approximation to lowest order \cite{Holstein:1940a}
\begin{equation} \label{eq:HP_firstorder}
	\begin{split} 
		\Fpx &\approx \sqrt{2F}a\\
		\Fmx &\approx \sqrt{2F}a^{\dag}\\
		\Fx &\approx F ,
	\end{split}
\end{equation}
where $\hat{F}_{\pm,x} = \Fy \pm i\Fz$, and $a,a^{\dag}$ are bosonic creation and annihilation operators.  We then write our state as $\ket{F,+F_x} = \ket{0}$, which is the vacuum in the Holstein-Primakoff representation.  Under this approximation, we can use the relations
\begin{align}
    \Sqt^{\dag}\Fx\Sqt &= F\\
    \Sqt^{\dag}\Fy\Sqt &= \frac{\sqrt{2F}}{2}e^{4F\xi_t}(a + a^{\dag})\\
    \Sqt^{\dag}\Fz\Sqt &= -i\frac{\sqrt{2F}}{2}e^{-4F\xi_t}(a - a^{\dag})   
\end{align}
to evaluate the expectation in Eq.\ (\ref{eq:hp_expectations}).  In light of this approximation, the tangent vector overlaps are readily shown to be
\begin{align}
     \expect{v_{\theta_t},v_{\theta_t}} &= \frac{Fe^{8F\xi_t}}{2}\\
     \expect{v_{\xi_t},v_{\xi_t}} &= 8F^2\\
     \expect{v_{\xi_t},v_{\theta_t}} &= 0 ,
\end{align}
where the last result indicates the tangent vectors are orthogonal as desired.

\subsubsection{Orthogonal Projection of Double-pass Filter}
Before performing orthogonal projection of the dynamics onto the tangent space, we must first convert the filtering equation from It\^{o} to Stratonovich form.  As is discussed in Ref.\ \cite{vanHandel:2005a}, the It\^{o} chain rule is incompatible with the differential geometry picture of projecting onto the tangent space.  Fortunately, Stratonovich stochastic integrals follow the standard chain rule and are thus amenable to projection filtering methods.  Following the derivation in  Appendix \ref{sec:app:ito_to_stratonovich}, we find that the Stratonovich SSE is given by
\begin{equation} \label{eq:double_pass_SSE_stratonovich}
	\begin{split} 
	    d\ket{\psi}_t &= \left[ -i\gamma B\Fy 
	        -M\left[ (\Fz - \expect{\Fz}_t)^2 - \expect{\Delta\Fz^2}_t\right]
		        -\frac{\sqrt{KM}}{2}\Fx \right.\\
		     & \left.
		        +2i\sqrt{KM}\expect{\Fz}_t\Fy
		        + i\sqrt{KM} \expect{\Fz\Fy}_t
		        \right]\ket{\psi}_t dt\\
		 &+ \left( \sqrt{M}(\Fz - \expect{\Fz}_t) + i\sqrt{K}\Fy\right)\ket{\psi}_t \circ dW_t ,
	\end{split}  
\end{equation}
where $\expect{\Delta\Fz^2}_t = \expect{\Fz^2} - \expect{\Fz}^2$.

In order to find the projection filter, we compare the general projection formula in Eq.\ (\ref{eq:general_projection}) to the general dynamical equation for states in our chosen family, given by
\begin{equation} \label{eq:projlhs}
    d\ket{\xi_t,\theta_t} = v_{\xi_t}d\xi_t + v_{\theta_t}d{\theta_t} .
\end{equation}
Using the orthogonality of the tangent vectors, the general forms for $d\xi_t$ and  $d\theta_t$ are
\begin{align}
    d\theta_t &= \frac{2e^{-8F\xi_t}}{F} \expect{v_{\theta_t}, d\ket{\psi_t}[\xi_t,\theta_t]}\\
    d\xi_t &= \frac{1}{8F^2}\expect{v_{\xi_t}, d\ket{\psi_t}[\xi_t,\theta_t]} ,
\end{align}
where $d\ket{\psi_t}[\xi_t,\theta_t]$ is the evolution of $\ket{\xi_t,\theta_t}$ under the Stratonovich filter of Eq.\ (\ref{eq:double_pass_SSE_stratonovich}).  

As an example calculation using these methods, consider projecting the dynamics generated by the magnetic field term.  Its contribution $c_\theta$ to the $\theta_t$ dynamics is given by
\begin{eqnarray}
   c_\theta & = &\frac{2e^{-8F\xi_t}}{F}\langle v_{\theta_t}, -i\gamma B \Fy \ket{\theta_t,\xi_t}dt\rangle
   	\nonumber \\
    & =& \frac{2\gamma Be^{-8F\xi_t}}{F}
        \bra{0}\Sqt^{\dag}\Yt^{\dag}\Fy^2\Yt\Sqt\ket{0}dt\\
    & =& \gamma B \bra{0}(a+a^{\dag})^2\ket{0} dt \nonumber \\
    & =& \gamma B dt \nonumber . 
\end{eqnarray}
Similarly, the contribution to $\xi_t$ is
\begin{eqnarray}
  c_\xi & = &\frac{ 1 }{8F^2}  \langle v_{\xi_t}, -i\gamma B \Fy 
        \ket{\theta_t,\xi_t}dt\rangle \nonumber \\
    & = & \frac{\gamma B}{4F^2}\bra{0} \Sqt{^\dag}(\Fz\Fy + \Fy\Fz)\Yt^{\dag}
    \Fy\Yt\Sqt\ket{0}dt \nonumber \\
   &  \propto& \bra{0} a^3 + a^2a^{\dag} - {a^{\dag}}^2a - {a^{\dag}}^3\ket{0}dt\\
    & =& 0  \nonumber.
\end{eqnarray}
Chugging through the remaining terms in a similar fashion, we arrive at the full projection filter equations
\begin{eqnarray}
    d\theta_t & = & \gamma B dt  
        + \frac{\sqrt{KM}}{2}e^{-8F\xi_t}\sin{\theta_t}dt
        + 2F\sqrt{KM}\sin{\theta_t} dt \nonumber\\
   &&     -\left[\sqrt{M}e^{-8F\xi_t}\cos{\theta_t}
         +\sqrt{K}\right] \circ dW_t
\end{eqnarray}
and 
\begin{equation}
    d\xi_t = \frac{M}{4}e^{-8F\xi_t}\cos^2{\theta_t}dt .
\end{equation}
Converting back to It\^{o} form using Eq.\ (\ref{app:eq:ito_to_stratonovich}), we have
\begin{subequations}\label{eq:projected_filter_ito}
	\begin{eqnarray} 
		    d\theta_t &= &\left[B \gamma 
		               - \frac{M}{4}e^{-16F\xi_t}\sin(2\theta_t)
		               + 2F\sqrt{KM}\sin{\theta_t}\right]dt \nonumber \\
		              && -\left[\sqrt{M}e^{-8F\xi_t}\cos{\theta_t} + \sqrt{K}\right]dW_t   \\
		    d\xi_t &= & \frac{M}{4}e^{-8F\xi_t}\cos^2{\theta_t}dt, 
	\end{eqnarray}
\end{subequations}
where the innovations are now in terms of the approximation of $\expect{\Fz}_t$ within the Gaussian family:
\begin{eqnarray}
	dW_t & = & dZ_t - 2\sqrt{M}\expect{\Fz}_tdt \nonumber \\
	& = & dZ_t + 2F\sqrt{M}\sin{\theta_t}dt .
\end{eqnarray}
\subsubsection{Small-angle Kalman Filter}
We see that the projected filter in Eq.\ (\ref{eq:projected_filter_ito}) is actually more general than the filters usually derived for the magnetometry problem, which do not distinguish the Gaussian and small-angle approximations.  That is, the family of states in Eq.\ (\ref{eq:gaussian_state_parameterization}) and the approximations considered in the above derivation only enforce the Gaussian state assumption through the Holstein-Primakoff approximation.  We can separately apply the small-angle approximation to recover an equation appropriate for the Kalman filter.  In this limit, the equation for $\xi_t$ completely decouples and has a closed form solution
\begin{equation}
    \xi_t = \frac{1}{8F}\ln\left[1+2FMt\right] .
\end{equation}
Taking the small-angle approximation for $\theta_t$ and plugging in the explicit form of $\xi_t$ gives
\begin{multline}
	   d\theta_t = \left[B\gamma + \left(2F\sqrt{KM} 
                - \frac{M}{2(1+2FMt)^2}\right)\theta_t\right]dt\\
                - \left[\frac{\sqrt{M}}{1+2FMt} + \sqrt{K}\right]dW_t, 
\end{multline}
which is linear in the remaining state parameter $\theta_t$.

While we could consider the Kalman filter for the quantum state alone, we can just as easily account for our uncertainty in $B$ at the same time.  That is, if we now embed $B$ as a state variable, setting $X_t = [ \theta_t\ B]^T$, the dynamics can be written in a linear form as
\begin{eqnarray}
    dX_t &=& A X_t dt + B dW_t \label{eq:linear_system}\\
    dZ_t &=& C X_t dt  + D dW_t \label{eq:linear_observations}\\
    A &=& \begin{pmatrix}
         2F\sqrt{KM} - \frac{M}{2(1+2FMt)^2} & \gamma\\
         0 & 0
    \end{pmatrix}\\
    B &=& \begin{pmatrix}
           -\frac{\sqrt{M}}{1+2FMt} - \sqrt{K}\\0
    \end{pmatrix}\\
    C &=& \begin{pmatrix}
        -2\sqrt{M}F & 0
    \end{pmatrix}\\
    D &=& 1.
\end{eqnarray}
Equations (\ref{eq:linear_system}) and (\ref{eq:linear_observations}) are precisely a classical linear system/observation pair, in which the same white noise process (the innovations) drives \emph{both} the system and observation processes.  The estimate $\tilde{X}_t = \mathbbm{E}[X_t | Z_{(0,t)}]$ admits a Kalman filter solution \cite{Liptser:1977}, given by
\begin{eqnarray}
    d\tilde{X}_t &= &A\tilde{X}_t dt + (B + VC^{\dag})d\tilde{W}_t\\
    \dot{V}    &= & AV +VA^{\dag} + BB^{\dag} - (B+VC^{\dag})(B+VC^{\dag})^{\dag} \nonumber
\end{eqnarray}
where $V$ is the covariance matrix
\begin{align}
    V &= \mathbbm{E}[(\tilde{X} - \mathbbm{E}[\tilde{X}])(\tilde{X} - \mathbbm{E}[\tilde{X}])^T]\\
      &= \begin{pmatrix}
        \Delta \tilde{\theta}_t^2 & \Delta\tilde{\theta}_t\tilde{B}_{kf}\\
        \Delta\tilde{\theta}_t\tilde{B}_{kf} & \Delta \tilde{B}_{kf}^2
      \end{pmatrix}
\end{align}
and
\begin{equation}
	d\tilde{W}_t = dZ_t + 2F\sqrt{M}\tilde{\theta}_t dt
\end{equation}
is the innovations constructed from the current $\theta_t$ estimate in the small-angle approximation.

\begin{figure*}[t]
    \centering
        \includegraphics[scale=1]{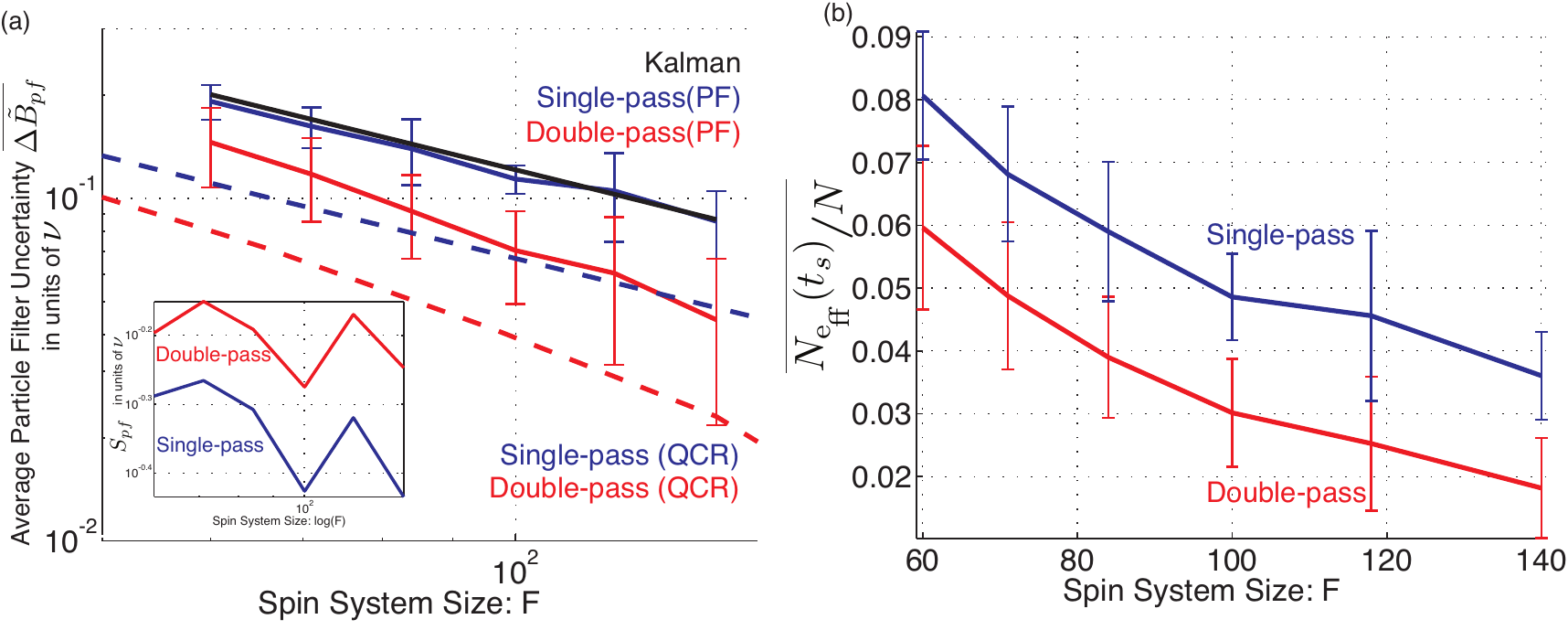}
    \caption{(color online) (a)  Estimator uncertainties as a function of $F$ averaged over 100 trajectories with $M = 10\nu$, $K=0.0006\nu$, $B=0$ and $t_s = 0.1\nu^{-1}$.  The initial $N = 1000$ particle set was drawn from a Gaussian distribution with mean zero and variance $10\nu^2$, which was also the same initial uncertainty in the Kalman filter $\Delta \tilde{B}_{kf}$.  A power-law fit to the particle filter (PF) scalings shows a single-pass scaling of $F^{-0.93}$ and a double-pass scaling of $F^{-1.39}$.  Also shown are the quantum Cram\'{e}r-Rao (QCR) bounds previously simulated for Figure \ref{fig:cramer_rao}. The inset shows the sample estimator deviation $S_{pf}$ for the same simulations. (b)  Average effective particle number fraction $\overline{N_\text{eff}(t_s)/N}$ as a function of $F$ for the data in (a).}
    \label{fig:particleFilterSimulations}
\end{figure*}

Looking at the explicit system of equations for the variances, we have
\begin{eqnarray} \label{eq:kalmanvar}
	   \frac{d(\Delta \tilde{\theta}_t^2)}{dt} &= & - M\Delta\tilde{\theta}_t^2
	     \left(\frac{1+4 F+8 F^2 M t}{(1+2 F M t)^2} 
	        + 4 F^2 \Delta\tilde{\theta}_t^2\right)\nonumber\\
	     & &+2 \gamma  \Delta\tilde{\theta}_t\tilde{B}_{kf}\\
	    \frac{d(\Delta \tilde{B}_{kf}^2)}{dt}&= &
	            -4F^2M (\Delta\tilde{\theta}_t\tilde{B}_{kf})^2\\
	        \frac{d(\Delta\tilde{\theta}_t\tilde{B}_{kf})}{dt} &= &
	           \gamma\Delta \tilde{B}_{kf}^2 -\frac{M}{2 (1+2 F M t)^2} 
	            \nonumber\\
	           & &  \left(1+4 F+8 F^2 M t+ \right. \\
	           && \left.  8 F^2 (1+2 F M t)^2
	               \Delta \tilde{\theta}_t^2\right)
	                \Delta\tilde{\theta}_t\tilde{B}_{kf} \nonumber
\end{eqnarray}
which are completely \emph{independent} of the second-pass coupling strength $K$.  That is, within the small-angle and Gaussian approximations, the double-pass system has no improvement in sensitivity and gives rise to the same $F^{-1}$ uncertainty scaling found previously for single-pass systems \cite{Geremia:2003a}. Perhaps this is unsurprising, as we attempted to find a linear description of an essentially non-linear affect.  Indeed, the numeric simulations in the next section suggest the single-mode Gaussian approximation breaks down just as the double-pass filter begins to show improved sensitivity to the magnetic field parameter.  Finally, we note that we have also derived a filtering equation which retains the next term in the Holstein-Primakoff expansion, but whose $K$ dependence nonetheless shows a negligible change relative to the lowest order expansion.

\section{Simulations} 
\label{sec:simulations}
Given the absence of an analytic improvement in the sensitivity of the quantum Kalman filter, we turn to numerical simulations of the quantum particle filter in order to gauge the potential of the double-pass system for magnetometry.  First recall how the filter would be used in an actual experiment.  Continuous measurements of the atomic cloud Larmor precessing under a particular, albeit unknown, magnetic field $B$ would give rise to the observations process $Z_{(0,t)}$.  This would then be fed into a classical computer to propagate the quantum particle filtering equations given in (\ref{eq:ensemble_filtering_eqs}).  The computer would then use the quantum particle set to provide the estimate $\tilde{B}_{pf}$ and uncertainty $\Delta \tilde{B}_{pf}^2$.  

In order to simulate such an experiment, we can generate the stochastic measurement record $Z_{(0,t)}$ using the quantum filter for the double-pass system given in Eq.\ (\ref{eq:double_pass_SSE}), evolved with a known magnetic field $B$.  Since the system is driven by the white noise process $dW_t$, the filtering equations may be integrated by the same integrator previously used to approximate the quantum Cram\'{e}r-Rao bound.  The measurements generated by these trajectories are equivalent to what the quantum particle filter would receive in an experiment, which means they can then be fed into the same particle filtering code to simulate an estimate of $B$.  In order to compare performance, we actually simulate two systems in parallel, one representing the double-pass system and the other, with $K=0$, representing a single-pass system.  Both utilize the same noise realizations on an individual trajectory.  

As is common when considering detection limits, we focus on the case of $B = 0$.  Although an unbiased estimator would assume no prior knowledge of the magnetic field value, such an approach is impractical for the particle filter, which would fail in approximating such large uncertainty with a finite number of particles.  As such, we take the initial distribution of $B$ values for the quantum particle set to be Gaussian
\begin{equation}
    p_0(B;\mu_B,\sigma_B) = \frac{1}{\sqrt{2\pi\sigma_B^2}}
    \exp(-\frac{(B-\mu_B)^2}{2\sigma_B^2})
\end{equation}
with mean $\mu_B = 0$ and variance $\sigma_B^2 = 10\nu^2$, where we again set $\gamma = 1$ and again define all parameters in units of $\nu$.  For a set of $N$ particles, the particle magnetic field values $\{B_i\}$ are drawn from the initial distribution, with weights $p_0^{(i)} = 1/N$.  The initial quantum state for all particles is set to the spin-coherent state along $+x$,i.e. $\ket{F,+F_x}$.  

Figure \ref{fig:particleFilterSimulations}(a) shows, with solid lines, the average particle filter uncertainty $\overline{\Delta \tilde{B}_{pf}}$ as a function of $F$, averaged over 100 measurement realizations using $N = 1000$ particles in each run of the filter. The error bars represent the the deviation in the simulated uncertainties over the 100 runs. As was the case for the Fisher information calculations, we observe an improved sensitivity scaling for the double-pass system, albeit with increased fluctuations in the individual run uncertainty $\Delta \tilde{B}_{pf}$.  Power-law least-squares fits of the average give a single-pass uncertainty scaling $F^{-0.93}$ and a double-pass scaling of $F^{-1.39}$ which are consistent with the quantum Cram\'{e}r-Rao scalings in figure \ref{fig:cramer_rao}.  Also shown is the analytic single-pass uncertainty scaling given by numerical integration of the Kalman covariance matrix via Eq.\ (\ref{eq:kalmanvar}).  We see that this agrees very well with the single-pass particle filter scaling and since it is consistent with previous Kalman filters used for magnetometry \cite{Geremia:2003a}, suggests the double-pass scaling does indicate improved sensitivity.  

\begin{figure*}[t]
\begin{center}
        \includegraphics[scale=0.4]{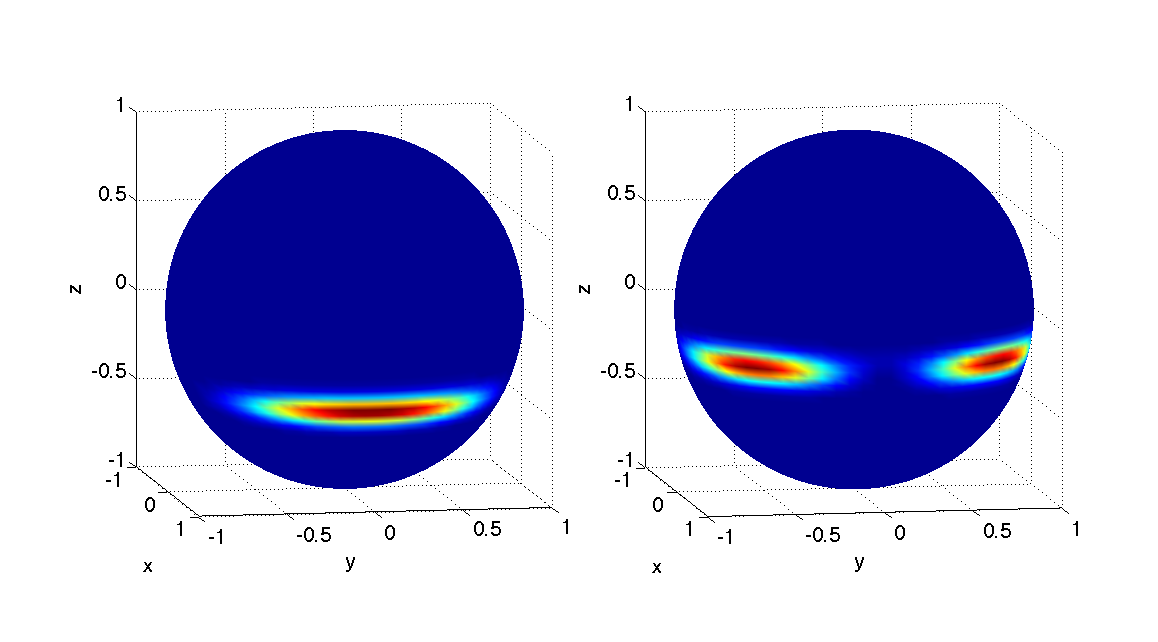}
\end{center}
\vspace{-10mm}
    \caption{(color online) Quasi-Probability distributions $Q(\theta,\phi,t)$ for two different trajectories at time $t_s = 0.1\nu^{-1}$ for $M = 10\nu, K = 0.0006\nu, B = 0$ and $F=140$.}
    \label{fig:splitting_example}
\end{figure*}

Of course, these statements are not without caveats.  The dashed lines in the plot correspond to the numerically computed quantum Cram\'{e}r-Rao bound, which is clearly below the estimates of all the filters.  This might mean that the continuous-measurement which gives rise to the numerical bound via Eq.\ (\ref{eq:app:quantum_filter_eq}) is simply not saturated by the corresponding estimator for that continuous-measurement.  Unfortunately, the above data took a week to generate on a quad-core workstation, indicating the technical challenges already present in simulating an $N=1000$ quantum particle set for the depicted range of $F$ limits the quality of the statistics.  As previously mentioned, the particle filter approximation is inherently biased, with the variance of estimates converging as $N^{-1}$.  
The inset in figure \ref{fig:particleFilterSimulations}(a) shows the sample estimator deviation $S_{pf}$, which is the deviation in the actual performance error of the particle filter on each individual run, i.e. $\tilde{B}_{pf} -B $ where the true $B = 0$.  In other words, $\Delta \tilde{B}_{pf}$ is the uncertainty calculated for an individual trajectory from the particle distribution $\{p_t^{(i)}\}$, which is averaged over many trajectories to get $\overline{\Delta \tilde{B}_{pf}}$.  However, an individual run of the particle filter also gives an estimate $\tilde{B}_{pf}$ of the true magnetic field $B$.  Since we know that the measurements were generated from a system evolved with $B = 0$, we can calculate the deviation in the actual estimates $\tilde{B}_{pf}$.  If the particle filter were unbiased, we would expect this sample deviation to equal the average particle filter deviation, i.e. $S_{pf} = \overline{\Delta \tilde{B}_{pf}}$.  Instead, the sample deviation dwarfs the average estimator uncertainty, indicating that the particle filter bias dominates.

We believe this is primarily a technical issue due to the dwindling contribution of initial particles to the final estimate.  Given that the potential particle magnetic field values $\{B_i\}$ are fixed at the outset, particles far from the true value will contribute only marginally to the final estimate.  That is, for $B_i$ far from the true $B = 0$, $p_{t_s}^{(i)}$ will be relatively small.  This behavior is well known in classical particle filtering \cite{Arulampalam:2002a} and is characterized by the effective sample size
\begin{equation}
    N_\text{eff}(t) = \frac{1}{\sum_{i=1}^N {p_t^{(i)}}^2}
\end{equation}
Figure \ref{fig:particleFilterSimulations}(b) shows the average effective particle number fraction $\overline{N_\text{eff}(t_s)/N}$ and its deviation as a function of $N$ for the same trajectories used in part (a) of the figure.  By the end of integration, we see that both single and double-pass systems have less than 100 of the initial 1000 particles significantly contributing to the estimate.  The fact that the double-pass system shows a smaller effective particle number is consistent with the increased sample estimator deviation $S_{pf}$.  In order to decrease the bias, one would need to include more particles or develop a resampling technique.  Nonetheless, we still believe the simulations depicted in figure \ref{fig:particleFilterSimulations}(a) suggest that the double-pass system shows in improved uncertainty scaling.

Numerical simulation also provides insight into how the Gaussian state assumption of the Kalman filter applies in the double-pass case.  Figure \ref{fig:splitting_example} shows quantum states evolved under two different noise realizations with $B=0,M=10\nu,K=0.0006\nu$.  Both states were initially spin-polarized along $+x$ and evolved until time $t_s = 0.1\nu^{-1}$ under the full double-pass SSE in Eq.\ (\ref{eq:double_pass_SSE}).  The Q-function shown is defined as
\begin{equation}
    Q(\theta,\phi,t) = \modsq{\braket{\theta,\phi}{\psi_t}}
\end{equation}
where the spin-coherent state $\ket{\theta,\phi}$ is the $+F$ eigenstate of the spin-operator $\Fx\sin\theta\cos\phi + \Fy\sin\theta\sin\phi + \Fz\cos\theta$.  Although one example shows a Gaussian squeezed spin state, the other shows a state with a bimodal Gaussian distribution, which we suspect is a consequence of the complicated nonlinear dynamics of Eq.\ (\ref{eq:double_pass_SSE})--- the presence of operators that are nonlinear in the spin observables, such as $\hat{F}_\mathrm{z} \hat{F}_\mathrm{y}$, do not preserve Gaussian states.  Therefore, a filter which confines the dynamics to the Gaussian family in Eq.\ (\ref{eq:gaussian_state_parameterization}) is only likely to remain valid for short time, and helps explain why the Kalman filter fails to find a difference between the single and double-pass setup.  These plots suggests a family of bimodal Gaussian states might result in a useful projection filter;  we have been unable to find a parameterization of such a family which admits an analytic derivation of a projection filter.

\section{Conclusions} 
\label{sec:conclusion}
We have explored the use of double-pass continuous measurement for precision magnetometry.  Our primary result involves numerical simulations of the quantum Cram\'{e}r-Rao bound which indicate that a double-pass system shows an improved magnetic field uncertainty scaling with atom number over a comparable single-pass system, albeit only for particular choices of coupling strengths relative to the collective spin size.  This is in contrast to quantum information theoretic bounds which suggest that the Heisenberg limit bounds the uncertainty scaling for both a single and double-pass system.  Clearly, future work aimed at reconciling these results is necessary, particularly deriving analytic quantum Cram\'{e}r-Rao bounds for unbounded ancilla systems.  However, at a heuristic level, we see that the double-pass provides a positive feedback effect which amplifies the amount of Larmor precession due to the magnetic field.  This of course also amplifies the initial spin-projection uncertainty, but the continuous measurement allows us to overcome this uncertainty by actually learning the state of the collective spin.  It is this interplay between amplification and measurement-induced squeezing which requires a fine-tuning of parameter strengths in order to find an improved Cram\'{e}r-Rao bound.

We have also explored estimators intended to achieve the uncertainty scaling seen in numerical simulations.  Taking a brute force approach, quantum particle filters show evidence of the improved double-pass scaling, although the results suffer from limited statistics which can not be significantly improved with current computational power and methods.  We are likewise limited to studying relatively small atomic ensemble sizes (small collective spin sizes), though we expect the improved scaling to extend to larger ensembles.  More practical quantum Kalman filters show no improved sensitivity, which is consistent with an observed breakdown in the Gaussian state assumption used to derive them.  However, the general projecting filtering technique used in the Kalman filter derivation provides an avenue for deriving more appropriate filters which might prove more tractable for practical magnetic field estimation.  More generally, we believe effective nonlinear interactions may prove an important tool in precision measurement.

We thank Rob Cook, Carl Caves, Anil Shaji, Steve Flammia, Sergio Boixo and Animesh Datta for many useful disagreements and discussions. This work was supported by the NSF (PHY-0639994) and the AFOSR (FA9550-06-01-0178).
\appendix
\section{Quantum It\^{o} Product Rule} \label{app:product_rule}
When manipulating products of variables governed by stochastic differential equations, one must be careful to use the It\^{o} product rule \cite{Oksendal:1946a}, which generalizes for the quantum case \cite{Bouten:2006a,Parthasarathy:1992a} as follows.  Consider two operators governed by the quantum stochastic differential equations
\begin{align}
    d\hat{X}_t &= \hat{A} d\St + \hat{B} d\St^{\dag} + \hat{C} dt\\
    d\hat{Y}_t &= \hat{D} d\St + \hat{E} d\St^{\dag} + \hat{F} dt    
\end{align}
then the product $\hat{X}_t\hat{Y}_t$ is governed by
\begin{align}
    d(\hat{X}_t\hat{Y}_t) &= \hat{X}_t(d\hat{Y}_t) + (d\hat{X}_t)\hat{Y}_t
                + (d\hat{X}_t)(d\hat{Y}_t) \nonumber\\
                &= (\hat{X}_t\hat{D} + \hat{A}\hat{Y}_t) d\St 
                    + (\hat{X}_t\hat{E} 
                        + \hat{B}\hat{Y}_t)d\St^{\dag}\nonumber\\
                &   + (\hat{X}_t\hat{F} + \hat{C}\hat{Y}_t + \hat{A}\hat{E}) dt   \label{eq:app:ito_product} ,
                    \end{align}  
where we have used the fundamental It\^{o} products---$d \St d \St^\dagger = dt$ and $d \St^\dagger d \St = d \St^2 = (d \St^\dagger)^2 = 0$.  Some find it useful to heuristically identify the white noise differentials $d\St,d \St^\dagger$ as order $\sqrt{dt}$, so that any consistent chain rule requires keeping terms to second order in differentials or equivalently, first order in $dt$.  

\section{Stochastic Schr\"{o}dinger Equation} \label{app:SSE}
Lacking any extra sources of decoherence, pure states remain pure under the dynamics described by the quantum filtering equation.  As such, it is often convenient for analysis and simulation to have a pure state description of the dynamics in terms of a \emph{stochastic Schr\"{o}dinger equation} (SSE).  Although these have previously appeared in the literature, for completeness we briefly derive the SSE for the general adjoint filter
\begin{align}
    d\rho_t &= -i[H,\rho_t]dt + \left(\hat{L}\rho_t\hat{L}^{\dag} - \frac{1}{2} \hat{L}^{\dag}\hat{L}\rho_t - \frac{1}{2}\rho_t\hat{L}^{\dag}\hat{L}\right)dt \nonumber\\
           & + \left( \hat{L}\rho_t + \rho_t \hat{L}^{\dag}
            - \Tr{(\hat{L} + \hat{L}^{\dag})\rho_t}\rho_t\right)dW_t .
\end{align}
We begin by writing
\begin{align}
	d\ket{\psi}_t &= A\ket{\psi}_tdt + B\ket{\psi}_tdW_t\\
	d\bra{\psi}_t &= \bra{\psi}_tA^{\dagger}dt + \bra{\psi}_tB^{\dagger}dW_t
\end{align}
From the It\^{o} rules, we have
\begin{eqnarray}
	d(\rho_t) &= & d(\ketbra{\psi}{\psi}_t) \nonumber \\
			 &= & \ket{\psi}_td(\bra{\psi}_t) + d(\ket{\psi}_t)\bra{\psi}_t + d(\ket{\psi_t})d(\bra{\psi}_t) \\
			  &= & (A\rho_t + \rho_tA^{\dagger})dt + (B\rho_t + \rho_tB^{\dagger})dW_t + B\rho_t B^{\dagger}dt \nonumber
\end{eqnarray}
Comparing the coefficients to the quantum filtering equation, we read off
\begin{align}
	B & = L - \expect{L}\\
	B^{\dagger} & = L^{\dagger} - \expect{L^{\dagger}}	
\end{align}
so that
\begin{equation}
	B\rho_tB^{\dagger} = L\rho_tL^{\dagger} - \expect{L^{\dagger}}L\rho_t
						 - \expect{L}\rho_tL^{\dagger}
						 + \expect{L}\expect{L^{\dagger}}\rho_t
\end{equation}
We try setting
\begin{equation}
	A = -\frac{1}{2}\left(L^{\dagger}L - 2\expect{L^{\dagger}}L 
				+ \expect{L}\expect{L^{\dagger}}\right)
\end{equation}
which means that
\begin{align}
	A \rho_t + \rho_t A^{\dagger} &=
					-\frac{1}{2}\left(L^{\dagger}L - 2\expect{L^{\dagger}}L
								+ \expect{L}\expect{L^{\dagger}}\right)\rho_t \nonumber\\
			&	  - \rho_t	\frac{1}{2}\left(L^{\dagger}L - 2\expect{L}L^{\dagger}
								+ \expect{L}\expect{L^{\dagger}}\right)\\
		&= -\frac{1}{2}L^{\dagger}L\rho_t -\frac{1}{2}\rho_t L^{\dagger}L
			+\expect{L^{\dagger}}L\rho_t \nonumber\\
		& + \rho_tL^{\dagger}\expect{L}
			- \expect{L}\expect{L^{\dagger}}\rho_t
\end{align}
and therefore
\begin{align}
	A \rho_t + \rho_t A^{\dagger} + B\rho_tB^{\dagger} 
			&= -\frac{1}{2}L^{\dagger}L\rho_t -\frac{1}{2}\rho_t L^{\dagger}L \nonumber\\
			&	+\expect{L^{\dagger}}L\rho_t + \rho_tL^{\dagger}\expect{L}\nonumber\\
			&	- \expect{L}\expect{L^{\dagger}}\rho_t
				+	L\rho_tL^{\dagger} \nonumber\\
			& - \expect{L^{\dagger}}L\rho_t
									 - \expect{L}\rho_tL^{\dagger}
									 + \expect{L}\expect{L^{\dagger}}\rho_t\\
			&= L \rho_t L^{\dagger} -\frac{1}{2}L^{\dagger}L\rho_t -\frac{1}{2}\rho_t L^{\dagger}L
\end{align}
which is the deterministic part of the quantum filtering equation as desired. 
\section{Converting between It\^{o} and Stratonovich SDE} 
\label{sec:app:ito_to_stratonovich}
Since white noise is non-differentiable, there are a variety of ways to give meaning to the differential $dW_t$ when integrated against some other stochastic process.  The most statistically pleasing is the It\^{o} integral, represented by the SDE
\begin{equation}
    dx_t = a(t,x_t)dt + b(t,x_t)dW_t 
\end{equation} 
and which satisfies $d\mathbbm{E}[x_t] = a(t,x_t)dt$, but requires using the It\^{o} chain rule.  Conversely, a Stratonivich SDE, written
\begin{equation}
    dx_t = \bar{a}(t,x_t)dt + b(t,x_t)\circ dW_t ,
\end{equation}
transforms using the normal chain rule but has a nontrivial expectation.  Fortunately, it is straightforward to convert between the two forms, which share the same stochastic coefficient and whose deterministic coefficients are related via
\begin{equation} \label{app:eq:ito_to_stratonovich}
    \bar{a}^{j}(t,x_t) = a^{j}(t,x_t) - \frac{1}{2} \sum_{k=1}^n b^{k}(t,x_t) \partialD{b^j(t,x_t)}{x_t^k}
\end{equation}
where the superscripts denote the $j$-th or $k$-th entry in the corresponding vector.  We recommend the interested reader consult \cite{Oksendal:1946a} for a more complete discussion of the various ways of constructing stochastic integrals and defining white noise increments.

For the double-pass It\^{o} SSE in Eq.\ (\ref{eq:double_pass_SSE}), we begin the conversion by noting that states with entirely real amplitudes form an invariant set and therefore write $\ket{\psi}_t = \sum_{m=-F}^F x_t^{m} \ket{m}$.  The stochastic coefficient is then 
\begin{equation}
	\begin{split}
	    b(t,x_t) &= \sqrt{M}\sum_{m = -F}^F m {x_t}^m\ket{m}
	                      - \sqrt{M}\sum_{m,n = -F}^F  n({x_t}^n)^2 {x_t}^m\ket{m}\\
	                      &+ \frac{1}{2}\sqrt{K}\sum_{m = -F}^F
	                        \left[ \sqrt{(F-m)(F+m+1)}{x_t}^m\ket{m+1} \right.\\
	                       & \left. - \sqrt{(F+m)(F-m+1)}{x_t}^m\ket{m-1}\right]
	\end{split}
\end{equation}
which has as its $j$-th entry
\begin{equation}
    \begin{split}
	    b^j(t,x_t) &= \sqrt{M} (j  - \sum_{n=-F}^{F}
	                                n({x_t}^n)^2){x_t}^j \\
	                &+\frac{\sqrt{K}}{2}
	                \left[\sqrt{(F-j+1)(F+j)}{x_t}^{j-1} \right.\\
	                & \left.
	                    - \sqrt{(F+j+1)(F-j)}{x_t}^{j+1}\right]
    \end{split}
\end{equation}
The derivative with respect to ${x_t}^k$ is then
\begin{equation}
	\begin{split}
	    \partialD{b^j(t,x_t)}{x_k} &= \sqrt{M} (j  - \sum_{n=-F}^{F}
	                                n({x_t}^n)^2)\delta_{jk}
	                                -\sqrt{M}2k{x_t}^k{x_t}^j \nonumber\\
	             &+\frac{\sqrt{K}}{2}
	                \left[\sqrt{(F-j+1)(F+j)}\delta_{(j-1),k} \right.\\
	             &\left. - \sqrt{(F+j+1)(F-j)}\delta_{(j+1),k}\right]
	\end{split}
\end{equation}
so that the sum in Eq.\ (\ref{app:eq:ito_to_stratonovich}) is
\begin{widetext}
\begin{eqnarray}
	    \sum_{k = -F}^Fb^k(t,x_t) \partialD{b^j(t,x_t)}{{x_t}^k} &= &
	             \sqrt{M} (j  - \sum_{n=-F}^{F}n({x_t}^n)^2)b^j(t,x_t)
	             - 2\sqrt{M}\sum_{k} k{x_t}^kb^k(t,x_t) {x_t}^j \\
	             & &+\frac{\sqrt{K}}{2}
	                 \left[\sqrt{(F-j+1)(F+j)}b^{j-1}(t,x_t)      - \sqrt{(F+j+1)(F-j)}b^{j+1}(t,x_t)\right] 
	                 	\nonumber
\end{eqnarray}
This suggests an equivalent operator form
\begin{eqnarray}    
	\left[\sqrt{M}(\Fz - \expect{\Fz}_t) + i\sqrt{K}\Fy\right]^2 
	- 2\sqrt{M}\expect{\Fz(\sqrt{M}(\Fz - \expect{\Fz}) + i\sqrt{K}\Fy)}_t
    & = & \left[\sqrt{M}(\Fz - \expect{\Fz}_t)+ i\sqrt{K}\Fy\right]^2 \\
    && - 2M\expect{\Delta\Fz^2}_t -2i\sqrt{KM}\expect{\Fz\Fy}_t \nonumber,
\end{eqnarray}
where $\expect{\Delta\Fz^2}_t = \expect{\Fz^2} - \expect{\Fz}^2$, so the Stratonovich SSE is
\begin{eqnarray}
    d\ket{\psi}_t &= &\left[ -i\gamma B\Fy -M\left[ (\Fz - \expect{\Fz}_t)^2 - \expect{\Delta\Fz^2}_t\right]  
    		-\frac{\sqrt{KM}}{2}\Fx
				        +2i\sqrt{KM}\expect{\Fz}_t\Fy
				        + i\sqrt{KM} \expect{\Fz\Fy}_t
				        \right]\ket{\psi}_t dt\nonumber\\
				 && + \left( \sqrt{M}(\Fz - \expect{\Fz})_t + i\sqrt{K}\Fy\right)\ket{\psi}_t \circ dW_t . \label{app:eq:SSE_stratonovich} 
\end{eqnarray}
\end{widetext}


\end{document}